%% file: EXO-24-005_temp.tex
\documentclass[11pt,twoside,a4paper,cmspaper,final,collab]{cms-tdr}
\def\svnVersion{2309a31-D}\def\svnDate{2026/08/18}\def\cmsCernNoTag{CERN-EP-2026-206}\def\cmsCernDate{\today}\def\cmsMessage{Submitted to the Journal of High Energy Physics}
\begin{document}\cmsNoteHeader{EXO-24-005}

\newcommand{\Mmuj}{\ensuremath{M_{\PGm\mathrm{j}}}\xspace}
\newcommand{\Wjets}{{\PW}+jets\xspace}
\newcommand{\DYjets}{DY+jets\xspace}
\newcommand{\ttjets}{{\ttbar}+jets\xspace}
\newcommand{\MLQ}{\ensuremath{M_{\text{LQ}}}\xspace}
\newcommand{\lumu}{\ensuremath{\lambda_{\PQu\PGm}}\xspace} 
\newcommand{\lbmu}{\ensuremath{\lambda_{\PQb\PGm}}\xspace} 
\newcommand{\LQumu}{\ensuremath{\mathrm{LQ}(\PQu\PGm)}\xspace}
\newcommand{\LQbmu}{\ensuremath{\mathrm{LQ}(\PQb\PGm)}\xspace} 
\newcommand{\metsig}{\ensuremath{\ptmiss/\sqrt{\sum E_{\text{T}}}}\xspace} 
\newcommand{\Nj}{\ensuremath{N_{\text{j}}}\xspace} 
\newcommand{\Nmu}{\ensuremath{N_{\PGm}}\xspace} 
\newcommand{\Ne}{\ensuremath{N_{\Pe}}\xspace} 
\newcommand{\Mmumu}{\ensuremath{M_{\PGm\PGm}}\xspace} 
\newlength{\cmsTabSkip}\setlength{\cmsTabSkip}{1ex}
\providecommand{\cmsTable}[1]{\resizebox{\textwidth}{!}{#1}}

\cmsNoteHeader{EXO-24-005}
\title{Search for scalar leptoquarks produced via muon-quark scattering in proton-proton collisions at \texorpdfstring{$\sqrt{s} = 13\TeV$}{sqrt(s) = 13 TeV}}

\date{\today}

\abstract{
 The first search for \TeV-scale scalar leptoquarks (LQs) produced in muon-quark ($\PGm\PQq$) interactions is presented. It is based on proton-proton collision data recorded at a center-of-mass energy of 13\TeV with the CMS detector at the LHC, corresponding to an integrated luminosity of 138\fbinv. Quantum fluctuations inside the proton generate charged-lepton components, enabling the study of lepton-induced processes. In particular, the interaction of a muon from one colliding proton with a quark from the other proton enables the study of resonant production of a single LQ and its subsequent decay. The final state includes one or two muons and a jet that can come from the hadronization of either a light quark (\PQu) or a \PQb quark. The primary observable is the invariant mass of the muon-jet system, whose distribution peaks at the LQ mass. The data are found to be in agreement with the background predictions. Upper limits are set on the product of the LQ production cross section and its decay branching fraction to the $\PGm\PQq$ final state. The results exclude broad regions in the parameter plane of the LQ mass and the LQ-$\PGm\PQq$ coupling, improving upon previous CMS searches by extending the coverage for LQs with masses between 1.5 and 3.6\TeV, and couplings between 0.2 and 0.6 for the \LQumu scenario; and masses above 1.8\TeV, and couplings above 1 for the \LQbmu scenario.
}

\hypersetup{%
pdfauthor={CMS Collaboration},%
pdftitle={Search for scalar leptoquarks produced via muon-quark scattering in proton-proton collisions at sqrt(s) = 13 TeV},%
pdfsubject={CMS},%
pdfkeywords={CMS, leptoquarks, muons}}

\maketitle 
\section{Introduction}
The standard model (SM) describes the interactions of elementary particles to high accuracy. However, there remain experimental observations in particle physics that are not fully explained, such as the similarities between 
the quark and lepton families. In some models of physics beyond the SM, these similarities are a manifestation of a deeper symmetry between these two types, and particles that bridge the two families naturally emerge. Leptoquarks (LQs) 
are hypothesized bosons that would manifest this fundamental connection between quarks and leptons and are predicted by numerous extensions of the SM, such as grand unified theories~\cite{PhysRevD.10.275,PhysRevLett.32.438,FRITZSCH1975193,PhysRevD.8.1240,Senjanovic:1982ex,PhysRevLett.64.619,PhysRevD.42.3892,Murayama:1991ah}, 
composite models with lepton and quark substructure \cite{SCHREMPP1985101}, technicolor models \cite{DIMOPOULOS1981370,DIMOPOULOS198069}, and superstring-inspired models \cite{HEWETT1989193}. Particles with similar decay 
modes are also found in $R$-parity-violating supersymmetry~\cite{FARRAR1978575,NILLES19841,BARBIER20051}. In addition, LQs have been proposed as a possible explanation of the anomalies observed in semileptonic \PB meson decays, 
including deviations in the lepton flavor universality ratios (such as $R_{\PDst}$)~\cite{PhysRevD.104.015007,Dorsner:2016wpm}, as well as, historically, in the anomalous magnetic moment of the muon $(g-2)$~\cite{PhysRevLett.116.141802,Du:2021zkq}, 
although the recently updated SM prediction is consistent with the measured value. These observations, in particular the flavor anomalies, provide further motivation to search for LQs coupled to muons and to first- or third-generation quarks.

Leptoquarks are hypothesized color-triplet scalar or vector bosons carrying both lepton and baryon numbers. They would decay either to a charged lepton and a quark, or to a neutrino and a quark.
In proton-proton ($\Pp\Pp$) collisions at the CERN LHC, LQs could be searched for via several experimental modes: pair production either via gluon fusion or quark-antiquark annihilation~\cite{Dorsner:2016wpm}; 
single production through quark-gluon fusion~\cite{Dorsner:2016wpm}; as a mediator of nonresonant production of lepton pairs in a $t$-channel scattering~\cite{PhysRevD.104.015007}; or by direct lepton-quark collisions~\cite{PhysRevLett.125.231804,greljo2020leptonquarkfusionhadroncolliders,Buonocore_2022}.
Analyses by the ATLAS and CMS Collaborations have predominantly concentrated on the first three production mechanisms~\cite{PhysRevD.109.112003,Pair_umu,CMS:2025iix}, as the lepton content of the proton was not known with sufficient precision to predict reliably the rate of the fourth process. Nevertheless, this production mode has recently been explored by the CMS experiment in final states including tau leptons~\cite{PhysRevLett.132.061801}. 
Advances in the determination of lepton and photon parton distribution functions (PDFs), particularly through the LUX formalism~\cite{Buonocore_2020,PhysRevLett.117.242002,Manohar_2017}, predict a sizable cross section 
for single LQ production via lepton-induced processes in $\Pp\Pp$ collisions~\cite{PhysRevLett.125.231804,greljo2020leptonquarkfusionhadroncolliders,Buonocore_2022}. In particular, for $\lambda \gtrsim 1$, where $\lambda$ 
denotes the Yukawa coupling in the LQ-lepton-quark vertex, the lepton-induced production cross section exceeds that of pair production by up to 2 orders of magnitude in the LQ mass (\MLQ) range of 1.5--5\TeV and remains 
the dominant single-production mechanism, while the quark-gluon-initiated contribution accounts for only 10--30\% of the total single-production cross section.

\begin{figure}[htbp]
   \includegraphics[width=0.48\textwidth]{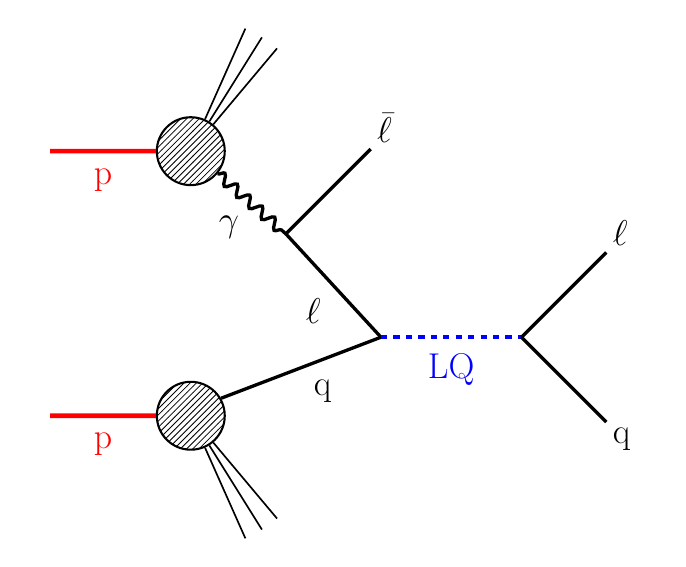}
   \centering
   \caption{An LO Feynman diagram of the lepton-induced LQ production at the LHC.}
   \label{fig:LQ_feynman}
\end{figure}

Lepton-induced LQ production proceeds via the fusion of a lepton and a quark, where the lepton is produced through a photon via quantum fluctuations in the proton. A leading order (LO) Feynman diagram for this process 
is shown in Fig.~\ref{fig:LQ_feynman}. We assume that a given LQ couples only to one lepton flavor. This assumption follows the minimal LQ benchmark models of
Refs.~\cite{Dorsner:2016wpm,PhysRevD.104.015007}, in which each LQ state has a single dominant coupling. Stringent constraints from lepton-flavor-violating processes and flavor-changing neutral currents require the 
coupling matrix to be strongly hierarchical, so that scenarios with comparable couplings to more than one lepton flavor are excluded for masses well beyond the range probed here. 
After the decay of the LQ, the final state consists of a high transverse momentum (\pt) lepton and a jet from the LQ decay, accompanied by a second same-flavor, oppositely charged lepton.
This second lepton is expected to have a smaller \pt than the primary lepton from the LQ decay and is likely to be produced outside the CMS detector coverage, as it is often emitted at small polar angles with respect to the beams.
This experimental signature differs from that of other production modes, whose final states typically contain two energetic leptons with similar \pt.

This analysis is based on $\Pp\Pp$ collision data at $\sqrt{s} = 13\TeV$ collected with the CMS detector in 2016--2018, corresponding to an integrated luminosity of 138\fbinv. Tabulated results are provided in the HEPData record for this analysis~\cite{hepdata}.

The search focuses only on the muon-induced production of scalar LQs that have a charge of $(1/3)\Pe$ or $(2/3)\Pe$, and can decay to a muon and an up quark (\PQu) or a bottom quark (\PQb), respectively; the two signal 
hypotheses are denoted as \LQumu and \LQbmu, and the associated Yukawa couplings by \lumu and \lbmu, respectively. The $\PQu\PGm$ coupling is chosen as the benchmark with the largest lepton-induced production cross section, 
while the $\PQb\PGm$ coupling is motivated by the LQ interpretation of the \PB meson anomalies. Scenarios with couplings to strange or charm quarks would yield intermediate sensitivities, bracketed by these two benchmarks. 
Other resonant LQ production mechanisms, such as pair production and gluon-induced single production, are not considered in this analysis. 
Pair production contributes negligibly in the considered phase space, while neglecting gluon-induced single production yields conservative sensitivity, as quantified in Section~\ref{sec:results}, with a negligible effect 
on the excluded (\MLQ, $\lambda$) region. Given that the search is optimized specifically for the lepton-induced production process, the results are presented considering only this mechanism, for simplicity.

\section{The CMS detector}
The central feature of the CMS apparatus is a superconducting solenoid of 6\unit{m} internal diameter, providing a magnetic field of 3.8\unit{T}. Within the solenoid volume are a silicon pixel and strip tracker, 
a lead tungstate crystal electromagnetic calorimeter (ECAL), and a brass and scintillator hadron calorimeter (HCAL), each composed of a barrel and two endcap sections. Forward calorimeters extend the pseudorapidity ($\eta$) coverage 
provided by the barrel and endcap detectors. Muons are measured in gas-ionization detectors embedded in the steel flux-return yoke outside the solenoid. More detailed descriptions of the CMS detector, together with a definition 
of the coordinate system used and the relevant kinematic variables, can be found in Refs.~\cite{CMS:2008xjf,CMS:2023gfb}.

Events of interest are selected using a two-tiered trigger system. The first level, composed of custom hardware processors, uses information from the calorimeters and muon detectors to select events at a rate of around 
100\unit{kHz} within a fixed latency of 4\mus~\cite{CMS:2020cmk}. The second level, known as the high-level trigger, consists of a farm of processors running a version of the full event reconstruction software optimized 
for fast processing, and reduces the event rate to around 1\unit{kHz} before data storage~\cite{CMS:2016ngn,CMS:2024aqx}.

\section{Event reconstruction}
The primary vertex is taken to be the vertex corresponding to the hardest scattering in the event, evaluated using tracking information alone, as described in Section~9.4.1 of Ref.~\cite{CMS-TDR-15-02}.
The global event reconstruction, also called particle-flow (PF) event reconstruction~\cite{CMS:2017yfk}, aims to reconstruct and identify each individual particle in an event, with an optimized combination of all subdetector 
information. In this process, the identification of the particle type (photon, electron, muon, charged hadron, neutral hadron) plays an important role in the determination of the particle direction and energy. Photons are 
identified as ECAL energy clusters not linked to the extrapolation of any charged-particle trajectory to the ECAL. Electrons are identified as a primary charged-particle track and potentially many ECAL energy clusters corresponding 
to this track extrapolation to the ECAL and to possible bremsstrahlung photons emitted along the way through the tracker material. Muons are identified as tracks in the central tracker consistent with either a track or several hits 
in the muon system, and associated with calorimeter deposits compatible with the muon hypothesis. Charged hadrons are identified as charged-particle tracks neither identified as electrons, nor as muons. Finally, neutral hadrons are 
identified as HCAL energy clusters not linked to any charged-hadron trajectory, or as a combined ECAL and HCAL energy excess with respect to the expected charged-hadron energy deposit.

For each event, hadronic jets are clustered from these reconstructed particles using the infrared- and collinear-safe anti-\kt algorithm~\cite{Cacciari:2008gp, Cacciari:2011ma} with a distance parameter of 0.4. Jet momentum 
is determined as the vectorial sum of all particle momenta in the jet, and is found from simulation to be, on average, within 5 to 10\% of the true momentum over the entire \pt spectrum and detector acceptance. Additional $\Pp\Pp$ 
interactions within the same or nearby bunch crossings (pileup) can contribute additional tracks and calorimetric energy depositions to the jet momentum. The pileup per particle identification algorithm (PUPPI)~\cite{CMS:2020ebo,Bertolini:2014bba} 
is used to mitigate the effect of pileup at the reconstructed particle level, making use of local shape information, event pileup properties, and tracking information. A local shape variable is defined, which distinguishes between 
collinear and soft diffuse distributions of other particles surrounding the particle under consideration. The former is attributed to particles originating from the hard scatter and the latter to particles originating from pileup 
interactions. Charged particles identified to be originating from pileup vertices are discarded. For each neutral particle, a local shape variable is computed using the surrounding charged particles compatible with the primary vertex 
within the tracker acceptance ($\abs{\eta} < 2.5$), and using both charged and neutral particles in the region outside of the tracker coverage. The momenta of the neutral particles are then rescaled according to their probability to 
originate from the primary interaction vertex, deduced from the local shape variable, superseding the need for jet-based pileup corrections~\cite{CMS:2020ebo}. Jet energy corrections are derived from simulation to bring the measured 
response of jets to that of particle-level jets on average. In situ measurements of the momentum balance in dijet, photon+jet, {\PZ}+jet, and multijet events are used to account for any residual differences in the jet energy scale 
between data and simulation~\cite{CMS:2016lmd}. The jet energy resolution amounts typically to 15--20\% at 30\GeV, 10\% at 100\GeV, and 5\% at 1\TeV~\cite{CMS:2016lmd}. Tight identification criteria are applied to each jet to remove 
jets potentially dominated by anomalous contributions from various subdetector components or reconstruction failures~\cite{CMS:2020ebo}.
Jets originating from bottom quarks are identified (``\PQb tagged'') using the \textsc{DeepJet} algorithm~\cite{Bols:2020bkb}, a multiclass jet flavor tagging algorithm that uses deep learning and exploits low-level features from a 
high number of jet constituents. The medium working point (WP) with a 75\% efficiency to identify \PQb quark jets and a 1\% misidentification probability for light-quark and gluon jets~\cite{Bols:2020bkb,CMS-DP-2018-058} is used.
Muons are measured in the range $\abs{\eta} < 2.4$, with detection planes made using three technologies: drift tubes, cathode strip chambers, and resistive-plate chambers. Matching muons to tracks measured in the silicon tracker results 
in a relative transverse momentum resolution, for muons with \pt up to 100\GeV, of 1\% in the barrel and 3\% in the endcaps. The \pt resolution in the barrel is better than 7\% for muons with \pt up to 1\TeV~\cite{CMS:2018rym}.

The missing transverse momentum vector \ptvecmiss is computed as the negative vector sum of the transverse momenta of all the PF candidates in the event, weighted by the PUPPI algorithm as described above, and its magnitude is denoted 
as \ptmiss~\cite{CMS:2019ctu}. The \ptvecmiss is modified to account for corrections to the energy scale of the reconstructed jets in the event. 

\section{Data and simulated samples}\label{sec:Data_and_MC}
The analysis uses a data set of $\Pp\Pp$ collisions at $\sqrt{s} = 13\TeV$ collected with the CMS detector during the 2016--2018 period, corresponding to an integrated luminosity of 138\fbinv~\cite{CMS-LUM-17-003,CMS-PAS-LUM-17-004,CMS-PAS-LUM-18-002}. 
The integrated luminosities for the 2016, 2017, and 2018 data-taking years have 1.2--2.5\% individual uncertainties~\cite{CMS-LUM-17-003,CMS-PAS-LUM-17-004,CMS-PAS-LUM-18-002}, while the overall uncertainty is 1.6\%. 
Events are selected using triggers that require at least one muon with $\pt > 50\GeV$, with no isolation requirements. The single-muon trigger efficiency exceeds 90\% over the full $\eta$ range.

Signal samples for scalar LQ production in the $s$-channel are generated at next-to-leading order (NLO) precision with \POWHEG v2 \cite{Nason_2004,Stefano_Frixione_2007,Alioli_2010,Jezo:2015aia} interfaced with 
\HERWIG 7.2.2~\cite{Bahr:2008pv,Bellm:2015jjp,Bellm:2019zci} for the showering, using the LUXlep lepton PDFs~\cite{Buonocore_2020}. The LUXlep PDF set corresponds to the combination of the NNPDF3.1luxQED set~\cite{Bertone:2017bme} 
and the lepton PDFs of Ref.~\cite{Buonocore_2020}. Two sets of signal samples are generated, for the $\PQu\PGm$ and $\PQb\PGm$ cases. The first set has \MLQ between 0.7 and 5.0\TeV and electric charge $(1/3)\Pe$, 
and the second has \MLQ in the 0.7--3.5\TeV range and electric charge $(2/3)\Pe$. In both cases, $\lambda$ is varied between 0.1 and 2. Charge-conjugate final states are simulated together. The intrinsic width of the 
LQ is computed consistently from $\lambda$ and \MLQ ($\Gamma_{\text{LQ}} = \lambda^{2} \MLQ/16\pi$), and remains small compared to the experimental mass resolution over the parameter space considered.

The main SM background processes in this analysis are \Wjets, \ttjets, and Drell--Yan (DY)+jets production. For the \Wjets and DY samples, the matrix element calculations include up to two additional outgoing partons. 
Other minor contributions from SM processes include diboson ($\PW\PW/\PW\PZ/\PZ\PZ$)+jets, associated \ttbar production ($\ttbar$+$\PZ/\PW/\PH$), and single-top quark production. Background from quantum chromodynamics (QCD) multijet processes is also considered.
The \DYjets, \Wjets, and diboson samples are generated at NLO using \MGvATNLO 2.6.5~\cite{Alwall_2014}, with FxFx jet merging applied to the \DYjets and \Wjets samples. The QCD multijet samples are generated with 
\PYTHIA 8.240~\cite{Sjostrand:2014zea}. Single-top quark and \ttbar samples are generated at NLO using \POWHEG. These generators are interfaced with \PYTHIA 8.240 for parton showering and fragmentation. The \PYTHIA parameters 
that affect the description of the underlying event are set according to the CP5 tune \cite{PITHYA8}. The PDF set used is NNPDF3.1~\cite{Ball_2017} at next-to-NLO precision. Simulated events include pileup, with a multiplicity 
distribution modeled to match the profile observed in data. The simulations are processed through a \GEANTfour~\cite{AGOSTINELLI2003250} model of the CMS detector.
An additional minor background (denoted ``Other'' in the plots) is represented by the lepton+jet final state from lepton-induced processes. This background process is generated using a configuration provided by the authors of Ref.~\cite{PhysRevLett.125.231804}, similar to that used for the LQ signal samples.
The simulated samples are corrected so that the detector response and resolution for both leptons and jets (including \PQb-tagged jets) and the trigger efficiency match those measured in data~\cite{CMS:2016lmd,CMS:2018rym}.

\section{Event selection}
Events are selected with at least one muon and at least one jet ($\Nj \geq 1$, where \Nj is the number of reconstructed jets) in the final state. We further divide the events into two independent signal regions (SRs), 
depending on the number of reconstructed muons ($\Nmu = 1$ or 2), as shown in Table~\ref{tab:Signal Region}. We refer to the reconstructed muon (jet) with the highest \pt in the event as the leading muon (jet), 
and to the muon with the second-highest \pt as the subleading muon. To reduce the overall background, while retaining most of the signal events, preselection criteria are applied to the recorded events, as well as to the 
simulated events that match the trigger requirements. This preselection requires one or two muons in the final state, depending on the SR. The leading muon must have $\pt > 55\GeV$, well above the trigger threshold so that 
the trigger efficiency is approximately constant, and the subleading muon (two-muon SR only) must have $\pt > 7\GeV$. All muons must be within $\abs{\eta_{\PGm}} < 2.4$ and satisfy high-\pt identification (ID) criteria 
and tracker isolation; the corresponding efficiencies are measured in data~\cite{CMS:2018rym}.
The leading jet is required to have $\pt>100\GeV$ and $\abs{\eta_{\text{j}}} < 2.4$. The invariant mass of the muon-jet system (\Mmuj) is required to be greater than 300\GeV, since LQ masses below the \TeV scale are already excluded by
previous searches~\cite{PhysRevD.109.112003,Pair_umu,CMS:2025iix}; this
requirement also strongly suppresses the SM background. To suppress top quark and diboson production events, we veto any events with one or more electrons ($\Ne = 0$, where \Ne is the number of reconstructed electrons). 
The quantity \metsig, where $\sum E_{\text{T}}$ is the scalar sum of the transverse energies of all PF candidates, computed consistently with the \ptmiss definition, must be less than $10\GeV^{1/2}$. 
Since the \ptmiss resolution scales approximately as $\sqrt{\sum E_{\text{T}}}$, the requirement on \metsig suppresses background events with genuine \ptmiss~\cite{CMS:2019ctu, MET_3}. The ratio of the \pt of the leading 
muon to that of the leading jet must be in the range $0.6<\pt^{\PGm}/\pt^{\text{j}}<1.6$ and the azimuthal angle between the leading muon and the leading jet ($\Delta\phi_{\PGm\text{j}}$) is required to be greater than 2.5. 
These requirements are applied to both SRs to reduce the SM background processes. In the two-muon SR only, the invariant mass of the two muons (\Mmumu) is required to be greater than 110\GeV, to remove the \DYjets background 
that peaks at the \PZ boson mass. The overall signal efficiency of the preselection, including the trigger, reconstruction, and
identification efficiencies as well as the kinematic requirements described above, is between 40 and 60\% for all signal hypotheses considered.

\begin{table}[ht]
   \topcaption{Preselections for the one- and two-muon SRs.}
   \centering
   \renewcommand{\arraystretch}{1.2}
   \begin{tabular}{l c c}
   Selection criteria & \multicolumn{2}{c}{Requirement} \\
   \hline
   Leading $\pt^{\PGm}$ & \multicolumn{2}{c}{${>}55\GeV$}\\
   $\abs{\eta_{\PGm}}$ & \multicolumn{2}{c}{${<}2.4$}\\
   Muon ID and isolation & \multicolumn{2}{c}{high-\pt ID, tracker isolation}\\
   Leading $\pt^{\text{j}}$ & \multicolumn{2}{c}{${>}100\GeV$}\\
   $\abs{\eta_{\text{j}}}$ & \multicolumn{2}{c}{${<}2.4$}\\
   \Mmuj & \multicolumn{2}{c}{${>}300\GeV$}\\
   \Ne & \multicolumn{2}{c}{${=}0$}\\
   $\pt^{\PGm}/\pt^{\text{j}}$ & \multicolumn{2}{c}{(0.6, 1.6)}\\
   $\Delta\phi_{\PGm\text{j}}$ & \multicolumn{2}{c}{${>}2.5$}\\
   \metsig & \multicolumn{2}{c}{${<}10\GeV^{1/2}$}\\
   \Nj & \multicolumn{2}{c}{${\geq}1$}\\[\cmsTabSkip]
    & One-muon SR & Two-muon SR \\
   \Nmu & ${=}1$ & ${=}2$ \\
   \Mmumu & \NA & ${>}110\GeV$ \\
   Subleading $\pt^{\PGm}$ & \NA & ${>}7\GeV$ \\
   \end{tabular}
   \label{tab:Signal Region}
\end{table}
The relevant backgrounds arise from SM processes with at least one jet and at least one muon in the final state. The most significant background contribution for the one-muon SR comes from \Wjets production, where the \PW boson 
decays leptonically. This scenario typically involves one lepton and large \ptmiss, and includes at least one jet that recoils against the \PW boson. Other background contributions also arise from lepton+jets \ttbar events, 
as well as from single-top quark and diboson production. The top quark decays into a \PW boson and a \PQb quark in these processes, yielding events with several high-\pt jets and, for leptonic \PW boson decays, a lepton. 
Another minor background process is QCD multijet production, where, in rare cases, high-\pt muons can be produced within a jet from decays of heavy mesons and be sufficiently isolated from other energy deposits.
For the two-muon SR, the main sources of backgrounds are processes with two prompt leptons, such as fully leptonic \ttbar decays and \DYjets production. These processes can also contribute to the one-muon SR when the second 
lepton fails the reconstruction or selection requirements.

\section{Signal optimization and final selection}
A boosted decision tree (BDT) is trained on background and signal events at preselection level to maximize the significance of a potential LQ signal in data. As inputs to the BDT, we select discriminant variables whose 
distributions exhibit minimal dependence on \MLQ, as the reconstructed mass \Mmuj will serve as the primary observable for the signal extraction. The simulated samples are split into two halves, one used for training and 
one for testing. The $\PQu\PGm$ signal used in the BDT training is a mixture of samples with different mass hypotheses. This choice is motivated by the fact that the input variables are designed to capture global kinematic 
features of the signal (event topology, energy scales, and lepton kinematic variables), which are largely insensitive to the quark flavor. Therefore, a training based on the $\PQu\PGm$ samples provides a robust and 
sufficiently general discriminant that can be applied to other signal hypotheses, including $\PQb\PGm$.
The background for the BDT training comprises a weighted mix of all the different SM simulated background samples described before. We apply correction factors that account for differences between data and simulation in the trigger, 
identification, and \PQb tagging efficiencies. The \textsc{AdaBoost} algorithm~\cite{FREUND1997119,Hocker:2007ht} is employed to train the BDT. To minimize correlations among input variables, only observables with low mutual 
correlations are included in the training. Independent assessments comparing BDT outputs on statistically independent training and testing event samples demonstrate consistent shapes and distributions. The compatibility 
of the two samples is quantified using the Kolmogorov--Smirnov test, as implemented in \textsc{tmva}~\cite{Hocker:2007ht}, yielding $p$-values that do not indicate any significant discrepancy, and therefore no evidence of overtraining.

The input variables for the BDT for the one-muon SR are the ratio of the \pt of the leading muon to that of the leading jet, the ratio of the \pt of the leading muon to the invariant mass of the muon-jet system, 
the azimuthal angle between the leading muon and the leading jet, the azimuthal angle between the leading muon and the \ptvecmiss, the \metsig quantity, the $\eta$ of the leading muon, the pseudorapidity difference 
between the leading muon and the leading jet, and the number of reconstructed jets in the event.
The BDT for the two-muon SR uses the same input variables, except for the azimuthal angle between the leading muon and the \ptvecmiss and the $\eta$ of the leading muon, which are removed, with the addition of the 
pseudorapidity difference between the leading and subleading muons, the ratio of the \pt of the subleading muon to the invariant mass of the muon-jet system, and the ratio of the \pt of the leading jet to the invariant mass of the muon-jet system.

\begin{figure}[htbp]
   \centering
   \includegraphics[width=0.48\textwidth]{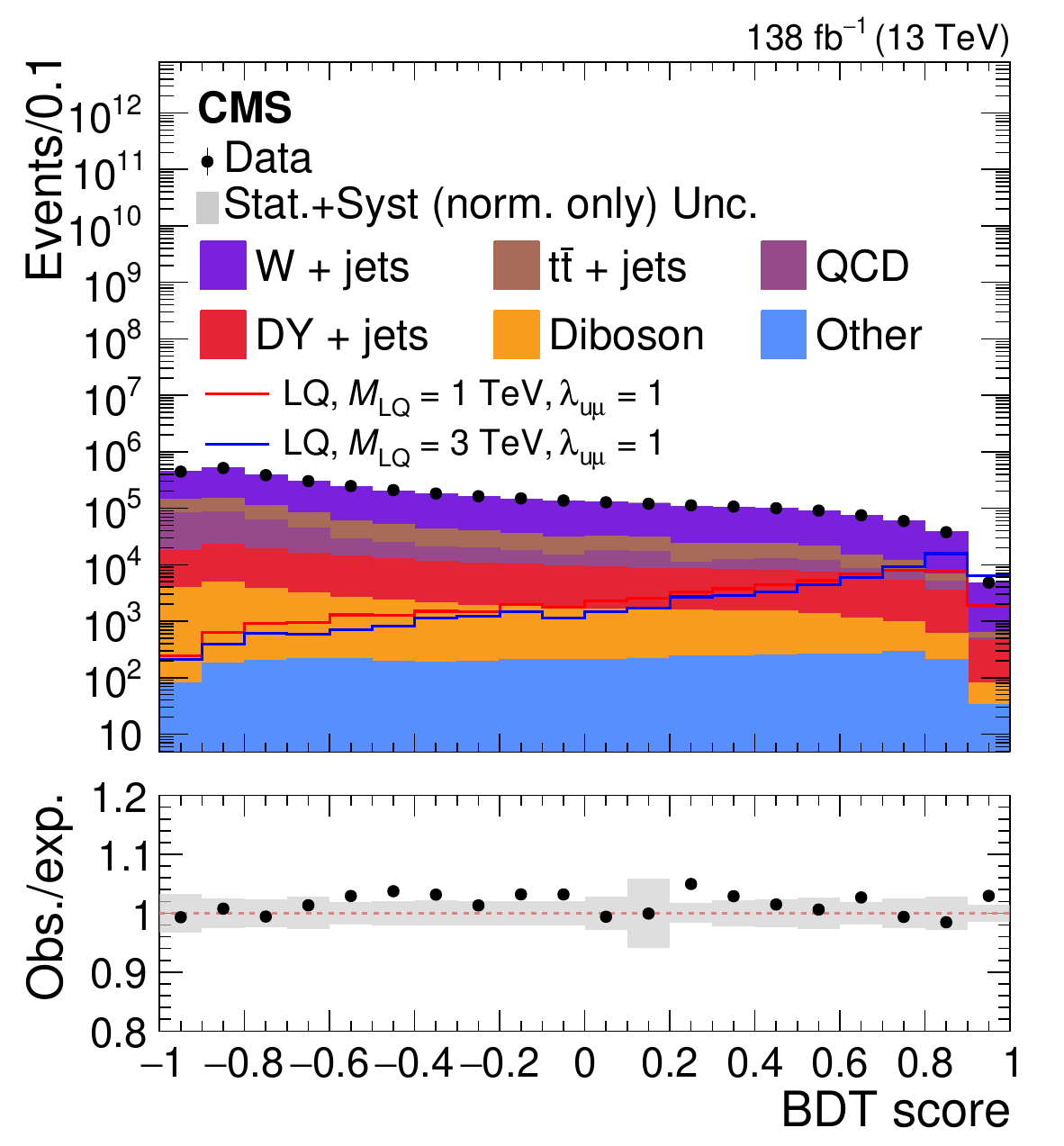}
   \includegraphics[width=0.48\textwidth]{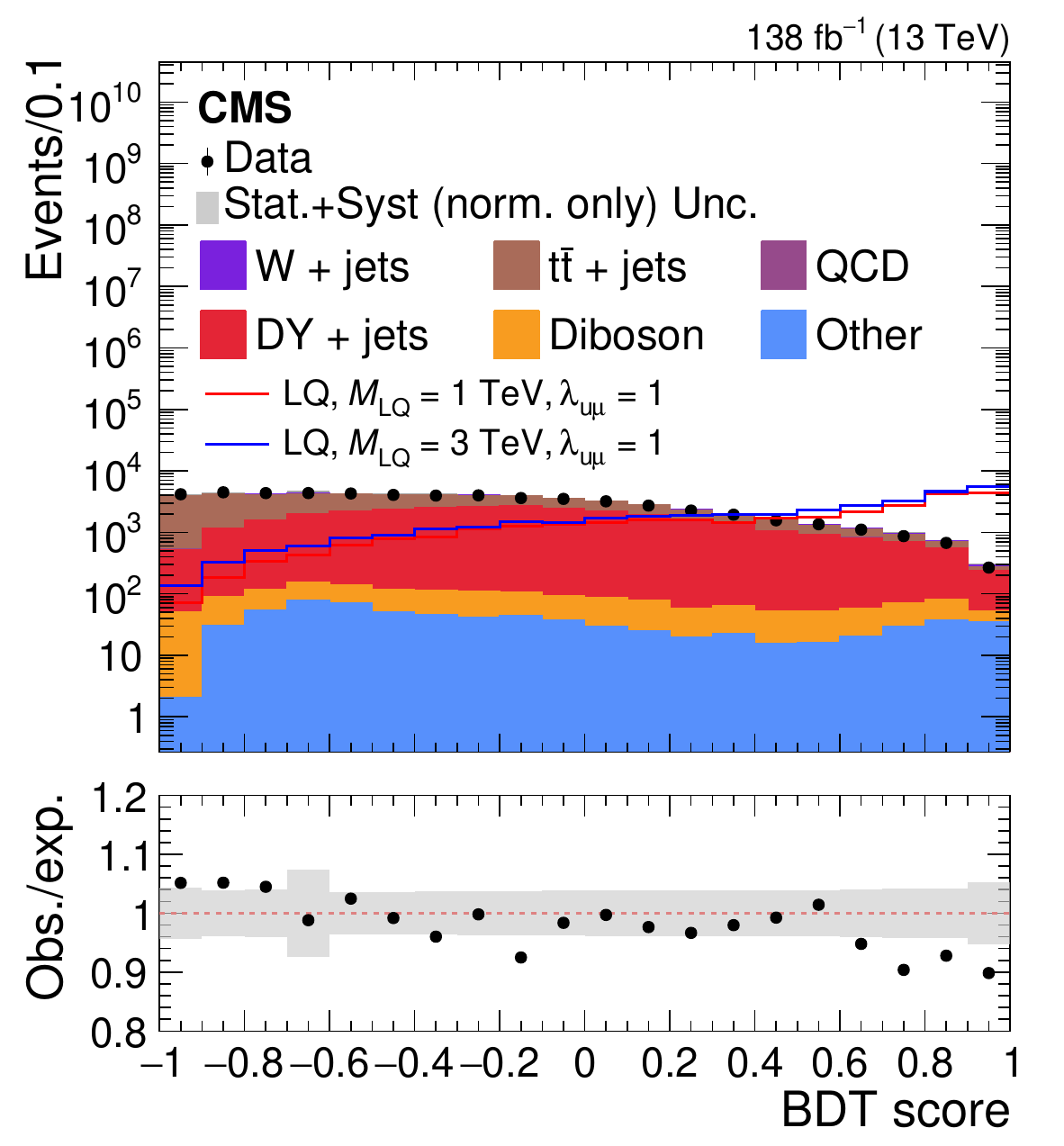}
   \caption{Comparison of data and background BDT discriminant distributions at the preselection level for the one-muon (left) and two-muon (right) SRs, shown for illustration purposes only. 
   The colored histograms show the simulated background prediction. The error bars are the data statistical uncertainties. The shaded band indicates an uncertainty of one standard deviation in the simulated background, 
   obtained by combining the statistical and normalization-only systematic contributions. The distributions of two simulated signal samples are also shown, corresponding to $\MLQ = 1$ or 3\TeV and $\lumu = 1$. 
   The signal cross section is set to 1\unit{pb} for visibility. The lower panels show the ratio of the data to the background prediction, with the shaded band indicating the same uncertainty as in the upper panels.}
   \label{fig:BDT_score}
\end{figure}

Comparisons of data and background BDT discriminant distributions at the preselection level for the one- and two-muon SRs are shown in Fig.~\ref{fig:BDT_score} (left) and (right), respectively. The separation of signal 
from background is optimized individually for both SRs for all \MLQ together using the relevant BDT discriminant as the sensitive variable. The figure of merit in the optimization is the Punzi significance~\cite{Punzi:2003bu} 
for a discovery potential of 3 standard deviations. This method is optimal both for making a discovery and for setting limits, and is also valid in cases with low background event counts.

For both SRs, we define two event categories based on the BDT score. The tighter category includes events with a BDT score greater than 0.4 (0.6) for the one-muon (two-muon) SR. This threshold is optimized to maximize 
signal significance across the range of signal hypotheses. The looser category includes events with lower BDT scores, from 0 to 0.4 and from $-0.2$ to 0.6 for the one- and two-muon SRs, respectively, recovering additional 
signal efficiency at the cost of lower purity. Events with scores below these ranges are excluded from further analysis because of their very low signal-to-background ratios. The optimal BDT boundaries do not vary with the coupling parameter $\lambda$.

After the BDT categorization, events are further categorized according to whether the leading jet, used to reconstruct \Mmuj, is \PQb tagged or not. The medium \textsc{DeepJet} WP, described above, is used.

In this way, four categories are defined in each SR, according to the loose or tight BDT requirement and the presence or absence of a \PQb tag on the leading jet. A summary of the resulting SR categories is presented in Table~\ref{tab:Categories}.

\begin{table}[htbp]
   \centering
   \topcaption{Final category selection for the SRs. The preselection requirements of Table~\ref{tab:Signal Region} are applied in all categories.}
   \cmsTable{
   \begin{tabular}{l c c c c c c c c}
    & \multicolumn{4}{c}{One-muon SR} & \multicolumn{4}{c}{Two-muon SR} \\ \hline
   \multirow{2}{*}{BDT score} & \multicolumn{2}{c}{$(0,0.4)$} & \multicolumn{2}{c}{${>}0.4$} & \multicolumn{2}{c}{$(-0.2,0.6)$} & \multicolumn{2}{c}{${>}0.6$} \\
    & \multicolumn{2}{c}{(loose)} & \multicolumn{2}{c}{(tight)} & \multicolumn{2}{c}{(loose)} & \multicolumn{2}{c}{(tight)} \\[\cmsTabSkip]
   Leading jet \PQb tag & No & Yes & No & Yes & No & Yes & No & Yes \\[\cmsTabSkip]
   Category name & 1$\PGm$0$\PQb$L & 1$\PGm$1$\PQb$L & 1$\PGm$0$\PQb$T & 1$\PGm$1$\PQb$T & 2$\PGm$0$\PQb$L & 2$\PGm$1$\PQb$L & 2$\PGm$0$\PQb$T & 2$\PGm$1$\PQb$T \\
   \end{tabular}
   }
   \label{tab:Categories}
\end{table}

\section{Background and signal models}
The targeted experimental signature is characterized by a localized excess on a smoothly falling \Mmuj spectrum shaped by contributions from multiple SM processes. This feature arises because of the resonant nature of 
the signal process. The signal strength is extracted from a maximum likelihood fit using all eight event categories considered simultaneously. The \Mmuj invariant mass distribution is binned using variable-width bins, 
chosen to match approximately the detector mass resolution across the spectrum. The fitting function for each category comprises two elements: a smoothly decreasing function $B$ to model the background and a signal 
histogram of the simulated \Mmuj spectrum, characterizing the resonance.

\subsection{Background estimation} \label{fit_background}
No significant variations are observed among the \Mmuj distributions corresponding to the three years of data taking, and a single fit to the combined data set is therefore performed. We model the SM background in each 
event category using an empirical function, also employed in previous CMS analyses~\cite{Sirunyan_2020_dij,2018_dijet,2013_EXO-11-094}, with the number of free parameters varying according to the specific category. The function
\begin{equation}
B(x) = p_0 (1-x)^{p_1} / x^{p_2} ,
\end{equation}
is used for all the categories, except for the 1$\PGm$1$\PQb$T category, which is fitted with the function
\begin{equation}
B(x) = p_0 (1-x)^{p_1} / x^{p_2+p_3\ln{x}}.
\end{equation}
In these functions, $x = \Mmuj/\sqrt{s}$, while the parameters $p_0$, $p_1$, $p_2$, and $p_3$ are treated as freely floating. The number of parameters chosen for each category is determined with a likelihood ratio (Fisher) test~\cite{fisher1922foundations,Wilks:1938dza}.

\begin{figure}[ht!] 
   \centering
      \includegraphics[width=0.48\textwidth]{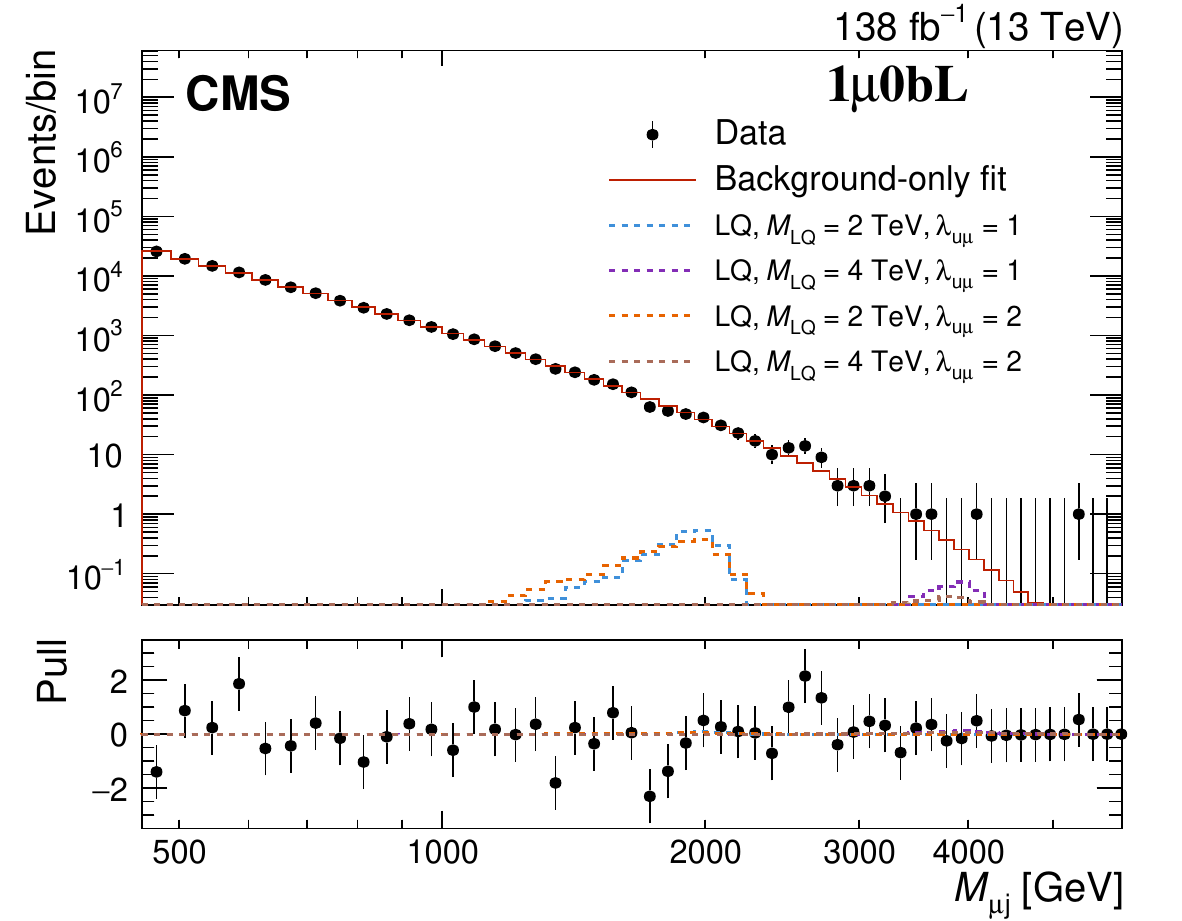}
      \includegraphics[width=0.48\textwidth]{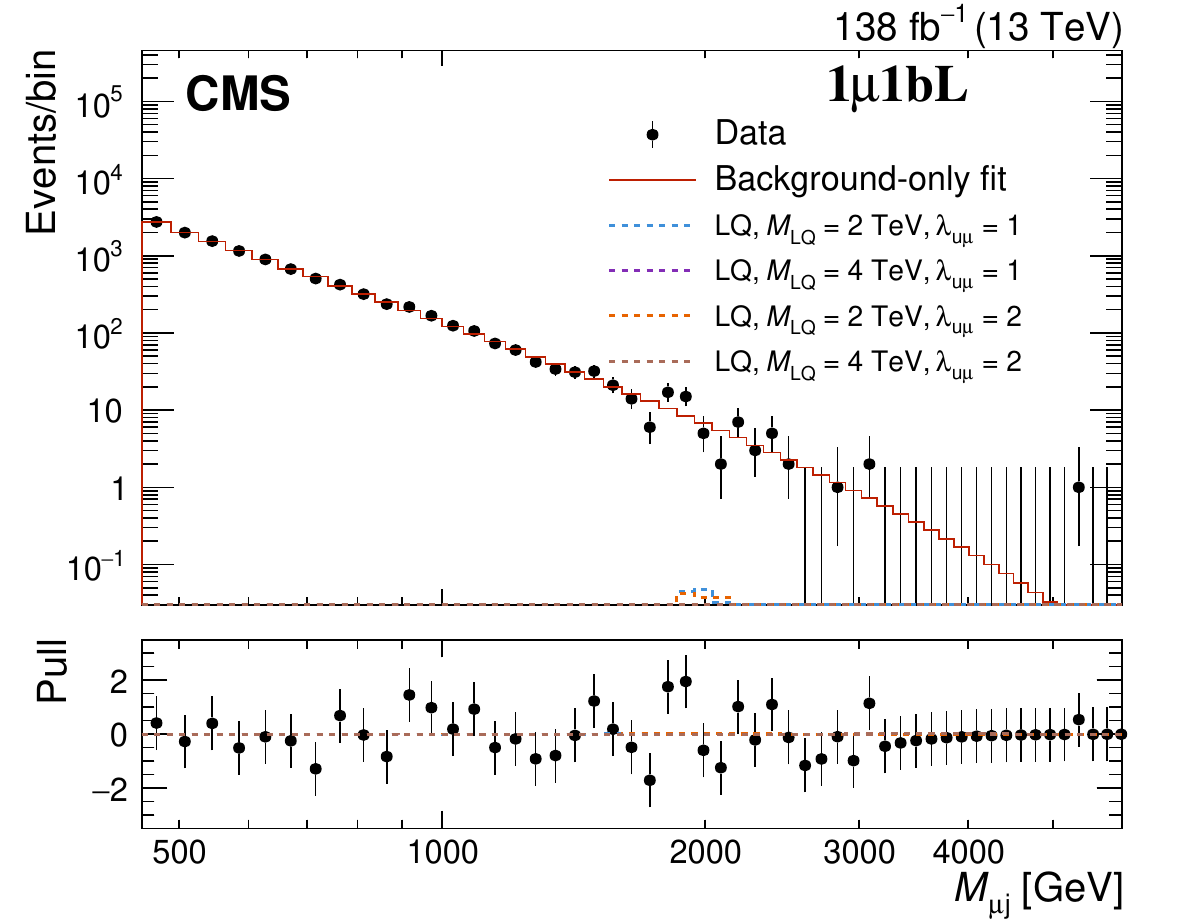}
      \includegraphics[width=0.48\textwidth]{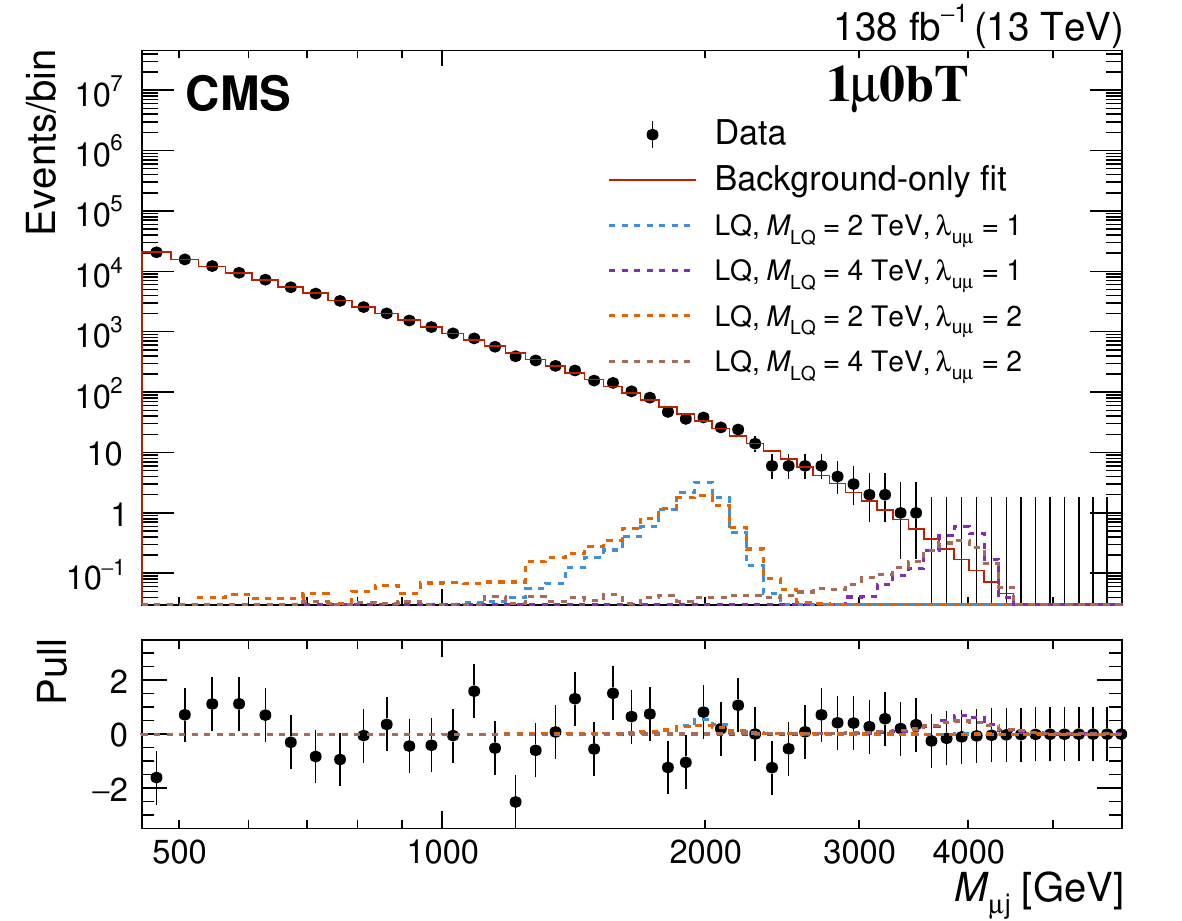}
      \includegraphics[width=0.48\textwidth]{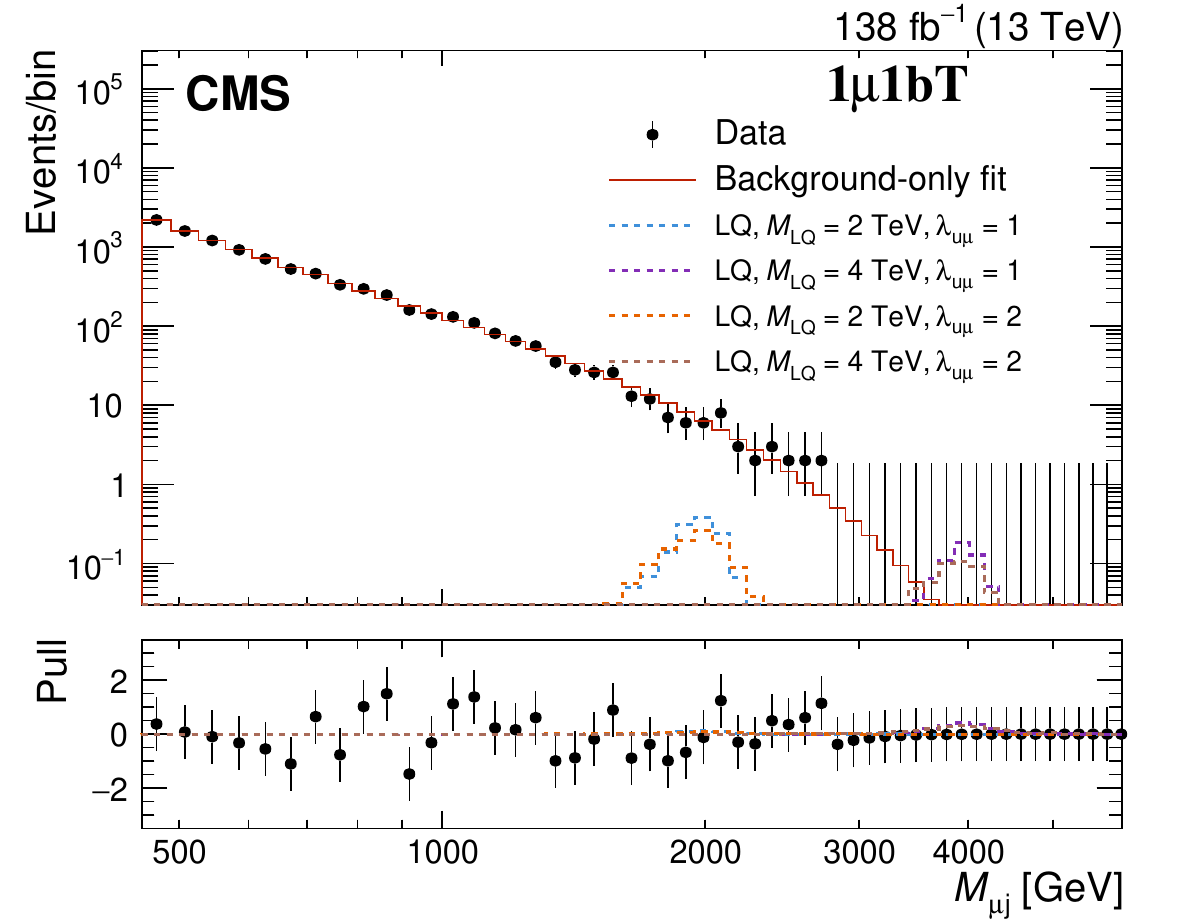}
      \caption{Empirical background function fits to the \Mmuj invariant mass distributions for all analysis categories within the one-muon SR: 1$\PGm$0$\PQb$L (upper left), 1$\PGm$1$\PQb$L (upper right), 1$\PGm$0$\PQb$T (lower left), 
      1$\PGm$1$\PQb$T (lower right). The upper panel in each frame shows the data distribution and the background component (red line) of the fit obtained using the empirical function. The data are shown as black points, 
      with vertical bars representing the statistical uncertainties computed as Poisson (Garwood) 68\% confidence intervals; in bins with no observed events only the upward interval is shown. The light blue (violet) 
      line represents a signal hypothesis with $\MLQ = 2$ (4)\TeV and $\lumu = 1$, while the orange (brown) line represents $\MLQ = 2$ (4)\TeV and $\lumu = 2$. All signals are normalized to the expected upper limit cross 
      section. The horizontal axis shows the \Mmuj value, while the vertical axis shows the number of events per bin. The lower panel in each plot shows the pulls for each bin, defined as $(N_{\text{data}} - N_{\text{fit}})/\sqrt{\smash[b]{N_{\text{data}} - \sigma_{\text{fit}}^2}}$, 
      where $N_{\text{data}}$ denotes the number of observed events, $N_{\text{fit}}$ the expected number of background events from the fit, and $\sigma_{\text{fit}}$ the uncertainty in $N_{\text{fit}}$.
      }
      \label{fig:background_estimation_dijet_1Muon}
   \end{figure}

\begin{figure}[ht!]
   \centering
      \includegraphics[width=0.48\textwidth]{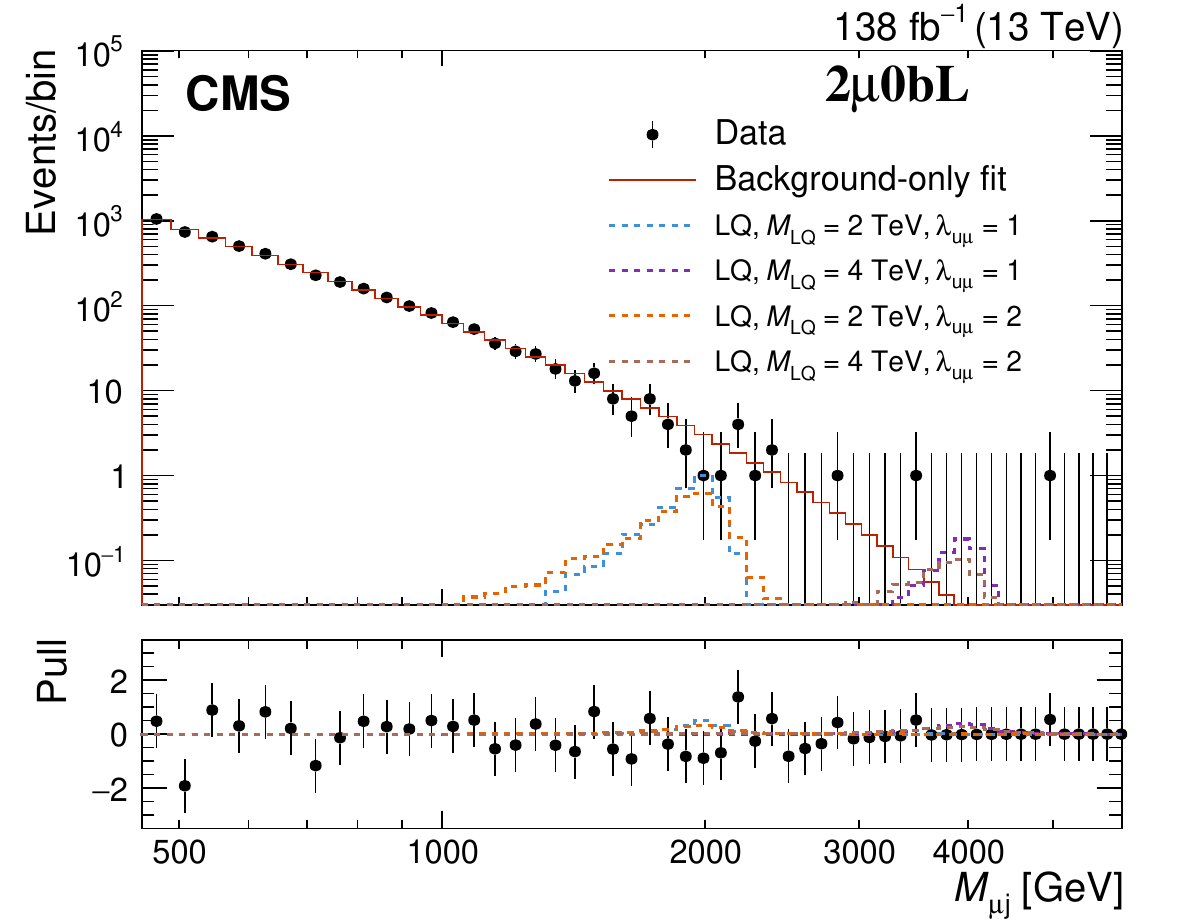}
      \includegraphics[width=0.48\textwidth]{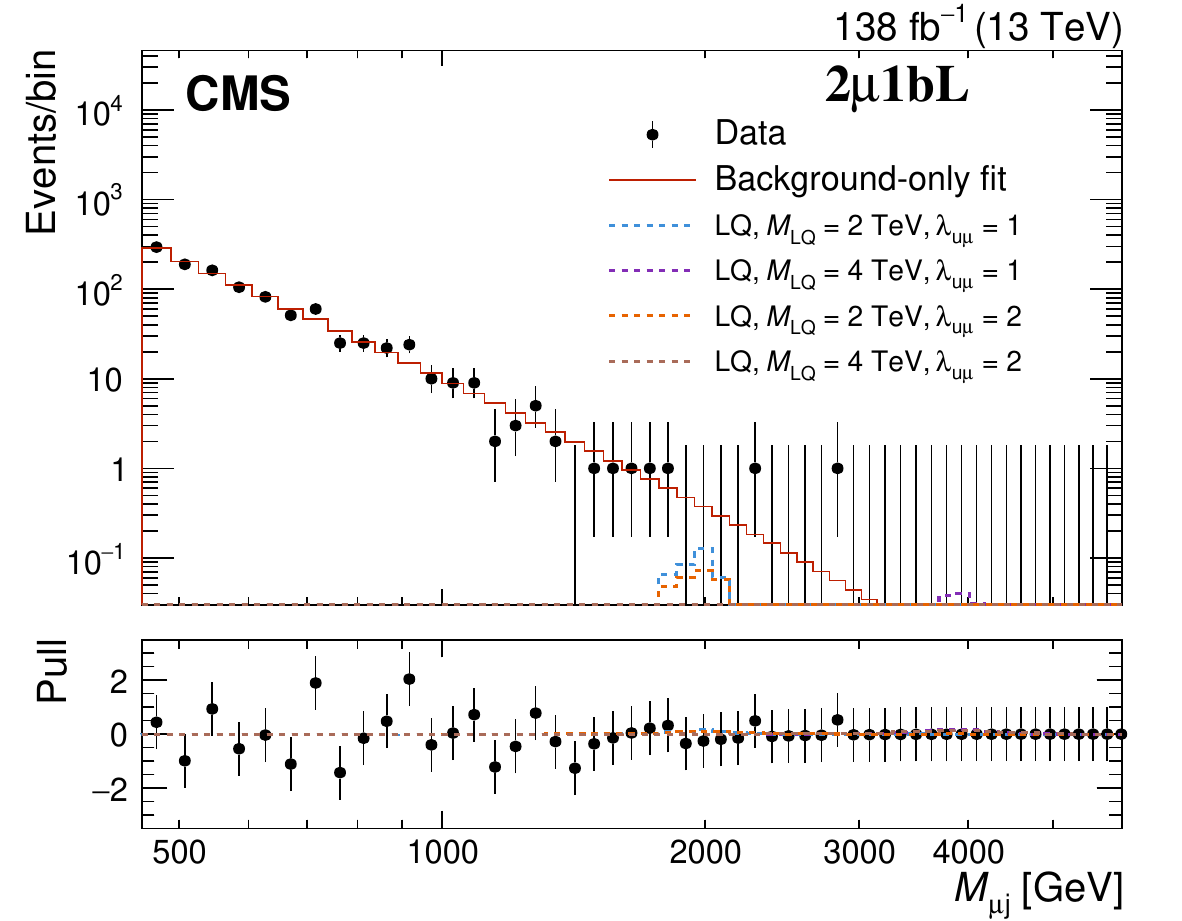}
      \includegraphics[width=0.48\textwidth]{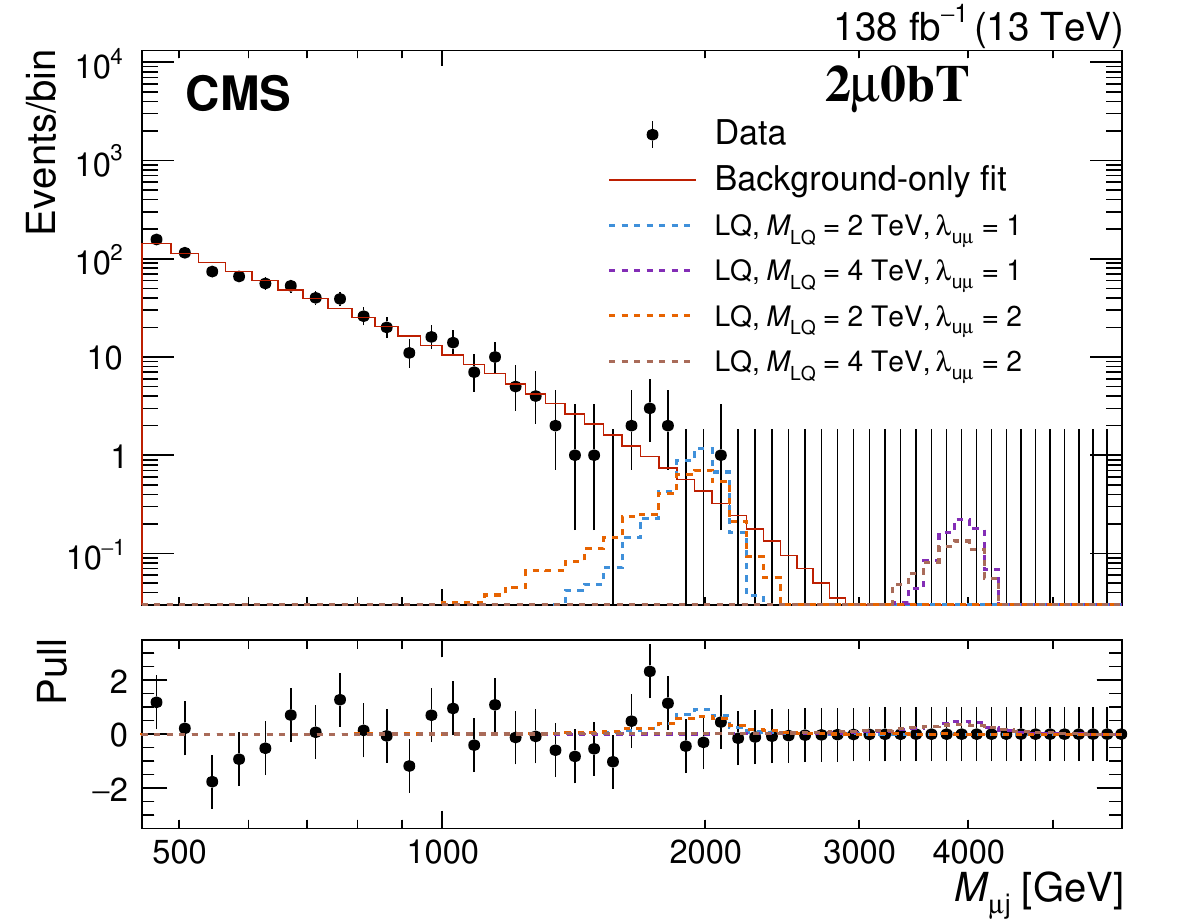}
      \includegraphics[width=0.48\textwidth]{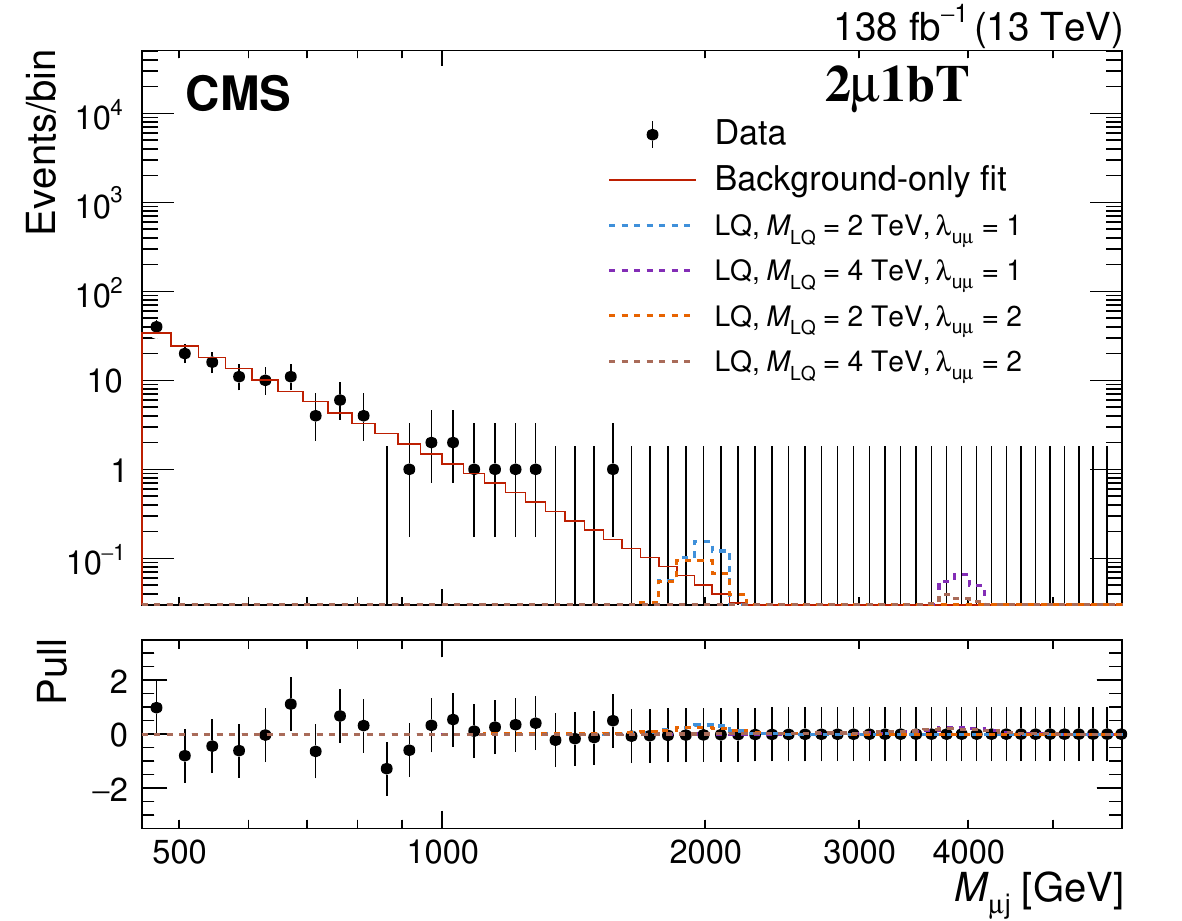}
      \caption{Empirical background function fits to the \Mmuj invariant mass distributions for all analysis categories within the two-muon SR: 2$\PGm$0$\PQb$L (upper left), 2$\PGm$1$\PQb$L (upper right), 2$\PGm$0$\PQb$T (lower left), 2$\PGm$1$\PQb$T (lower right). 
      The notations are as in Fig.~\ref{fig:background_estimation_dijet_1Muon}.
      }
      \label{fig:background_estimation_dijet_2Muon}
   \end{figure}

We choose a common lower bound of 450\GeV for the fit range in all categories to ensure consistency across the analysis. Extending the fit below this value introduces significant deviations from the smooth, monotonically decreasing behavior assumed for the background model, 
thus degrading the fit quality. Since the search targets resonances with a minimum invariant mass at the \TeV scale, the precise starting point of the fit is not critical for this analysis. The fits with the background-only 
function are shown in Figs.~\ref{fig:background_estimation_dijet_1Muon} and \ref{fig:background_estimation_dijet_2Muon}.

\subsection{Signal model}
Signal shapes are generated using the simulated samples described in Section~\ref{sec:Data_and_MC}. For each signal hypothesis defined by specific values of \MLQ and $\lambda$, a single shape normalized to unity is used 
for interpolation, independently for each event category. Signal shapes are first interpolated between the LQ mass points spaced by 100\GeV, at fixed values of the coupling parameter. In a second step, interpolation is 
performed between coupling values in steps of 0.25 at a fixed mass, accounting for the corresponding change in the reconstructed mass distribution, whose width increases with $\lambda$ ($\Gamma_{\text{LQ}} \propto \lambda^2$), 
leading to a broader resonance for larger coupling values. The interpolation is performed using the quantile function interpolation algorithm described in Ref.~\cite{READ1999357}.

In conjunction with signal shapes, the signal efficiencies across each category are interpolated for intermediate values of \MLQ and $\lambda$. The signal efficiency for each category is defined as $\epsilon_i = N_{i}/N_{\text{gen}}$, 
where $N_{i}$ is the number of events passing the final selection for the category $i$, and $N_{\text{gen}}$ is the total number of generated events for the considered signal sample. This efficiency is interpolated as a function of \MLQ 
and $\lambda$ with a polynomial function, whose order is chosen for each category as the lowest that describes the generated points.

\begin{figure}[ht!]
   \centering
   \includegraphics[width=0.48\textwidth]{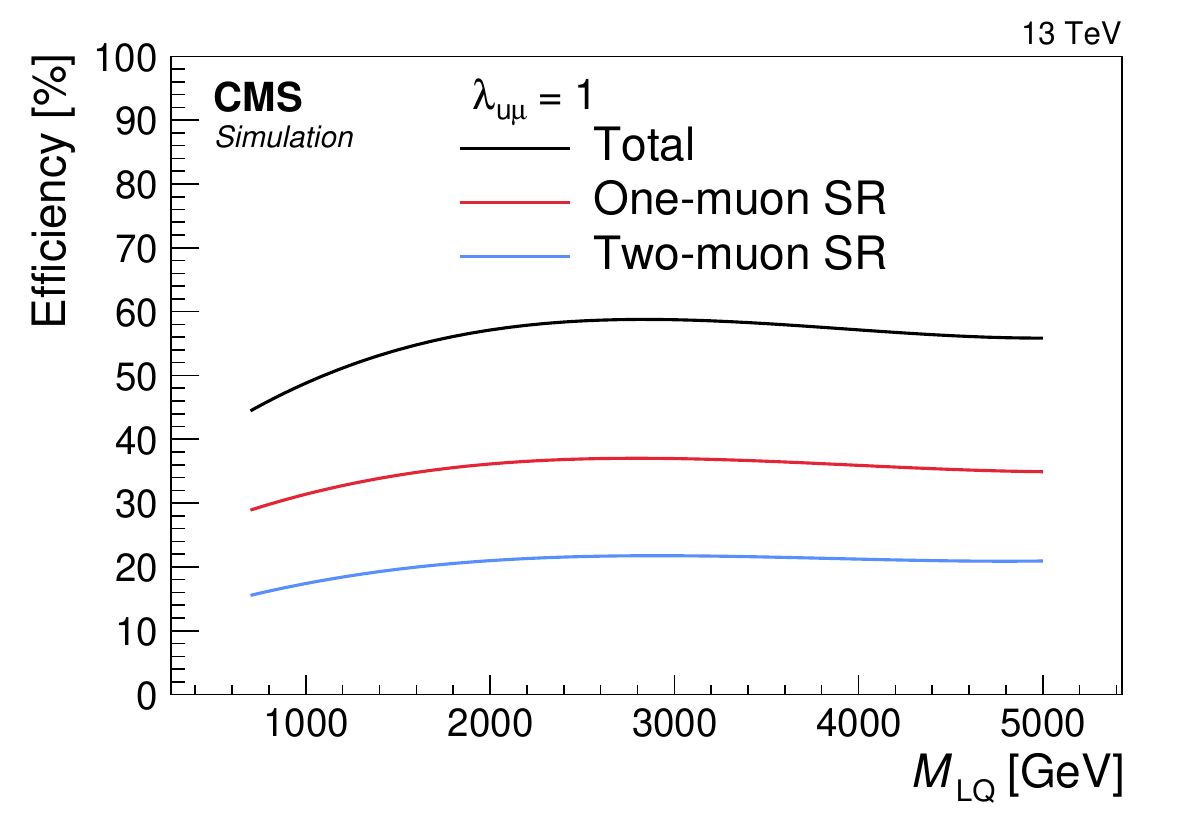}
   \includegraphics[width=0.48\textwidth]{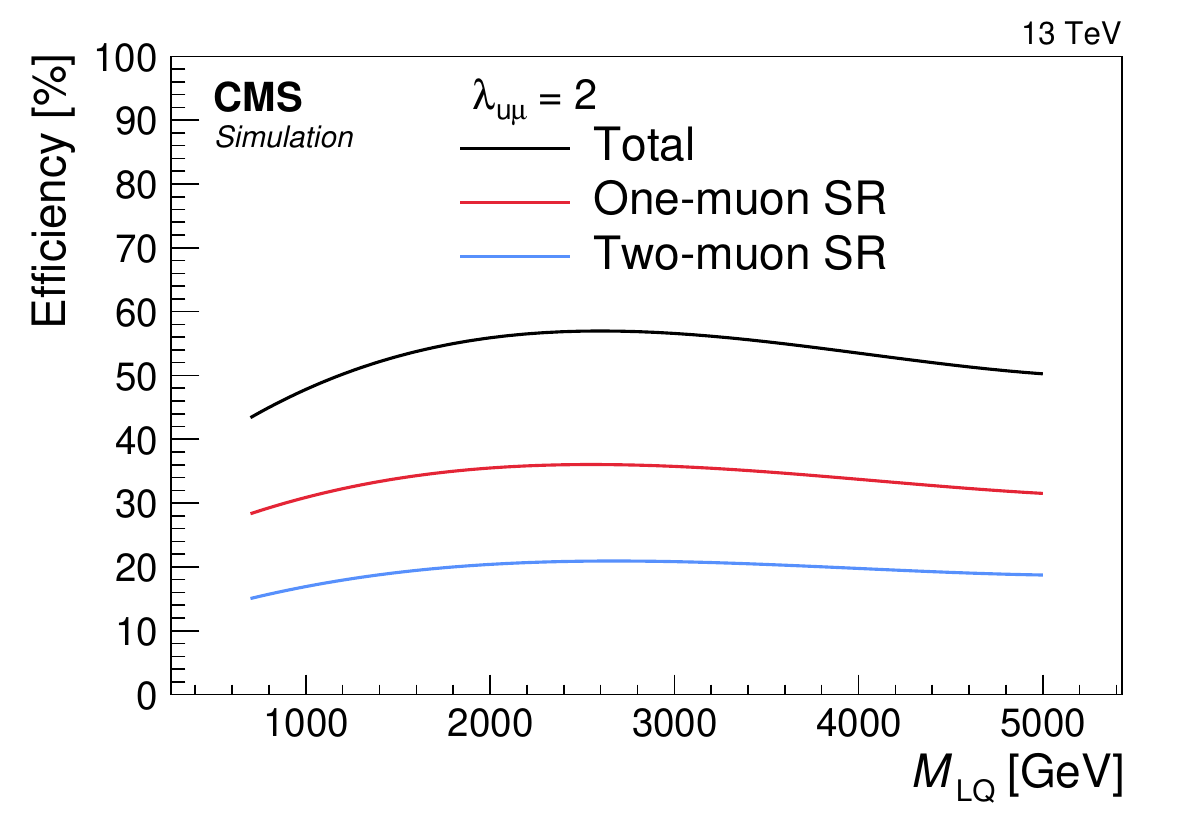} 
   \includegraphics[width=0.48\textwidth]{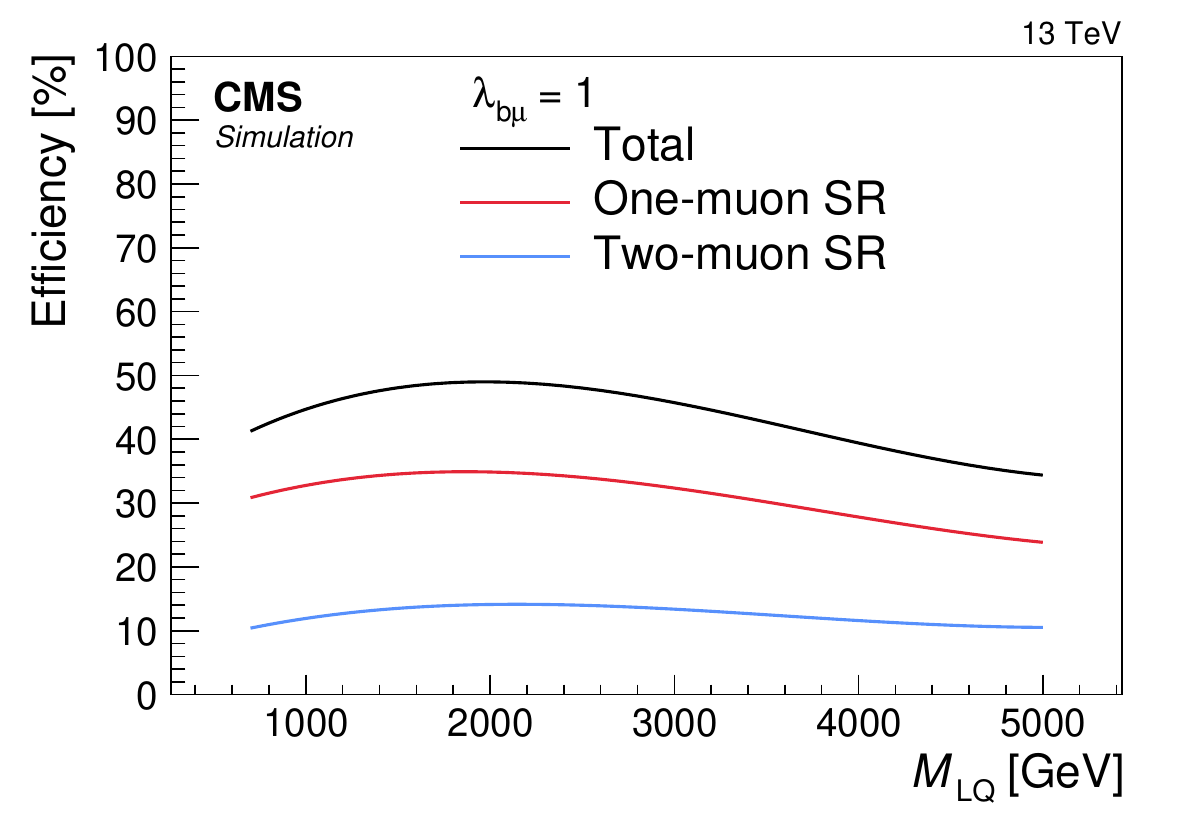}
   \includegraphics[width=0.48\textwidth]{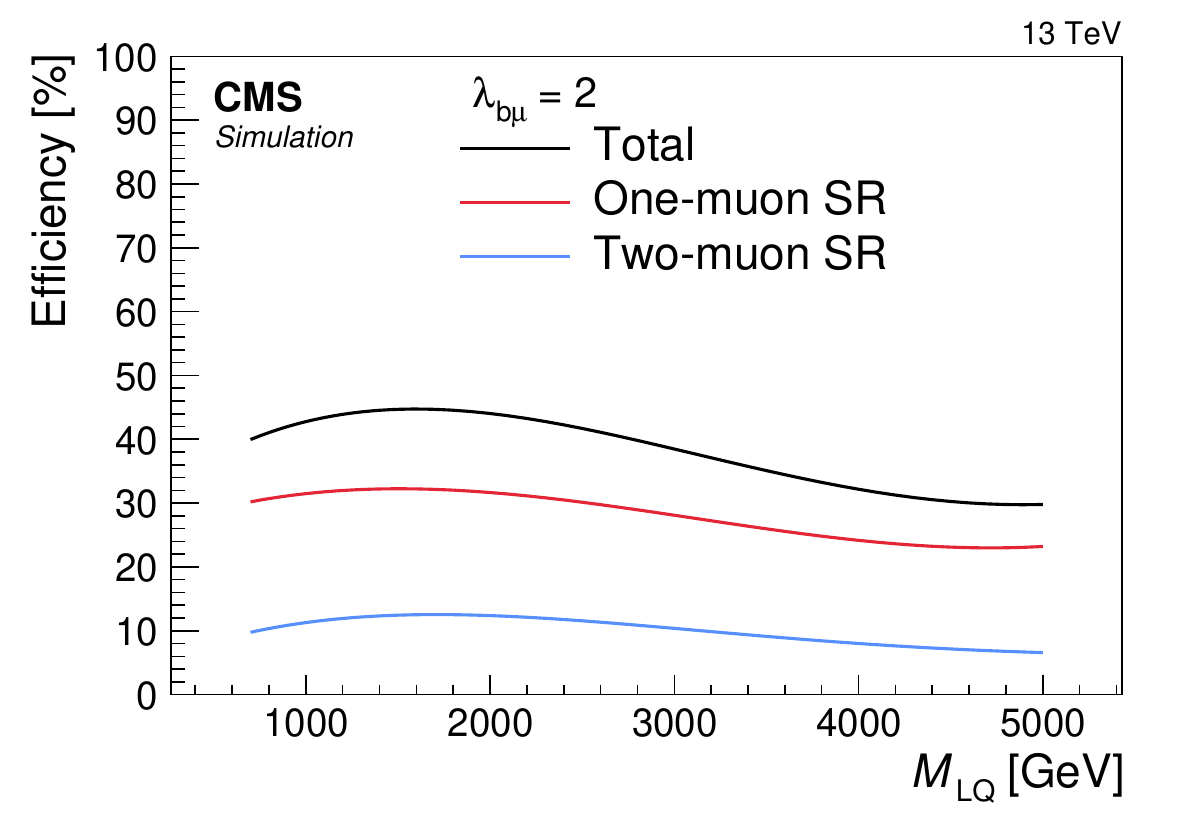}
   \caption{Signal efficiency as a function of \MLQ for $\lumu = 1$ (upper left), $\lumu = 2$ (upper right), $\lbmu = 1$ (lower left), and $\lbmu = 2$ (lower right). The black curves are the total efficiency, defined as the 
   sum of the efficiencies of the two SRs. The red and blue curves are the efficiency for the one- and two-muon SRs, respectively.}
   \label{fig:efficiency}
\end{figure}

The efficiency for each SR, as well as the total efficiency, are shown in Fig.~\ref{fig:efficiency} for the $\PQu\PGm$ and $\PQb\PGm$ final states. In the $\PQb\PGm$ case, the efficiency decreases at higher \MLQ because 
of the larger low-mass tail of the signal \Mmuj distribution, caused by the steeply falling \PQb quark PDF, which enhances off-shell production below the resonance pole.

\subsection{Fit validation}
Detailed signal injection tests are carried out to look for any potential bias in the background prediction method.   
The signal injection tests are performed as follows: we generate pseudo-data distributions for a hypothesis including background but no signal, using the background function, with parameters fixed to the values from the best fit to data. 
We also produce pseudo-data distributions including both the background and a signal, injected with a cross section equal to three times the expected cross section limit. We generate these distributions for all considered signal hypotheses. 
For each pseudo-data distribution, we repeat the fitting procedure and extract the corresponding fitted signal cross section and its standard deviation. We examine the distribution of the bias in units of standard deviations: that is, 
the difference between the fitted and the injected signal cross sections, divided by the standard deviation of the fit. For all resonance masses, widths, and signal strengths considered, the mean bias is below half a standard deviation, 
and in the vast majority of the cases it is even much smaller. In addition, these studies are performed with the pseudo-data distributions generated from an alternative empirical function
   $\exp(-p_2 p_0/x)/(p_0/x)^{p_1}$
that also describes the data. This tests the flexibility of the background function to fit to a spectrum with a different shape. The entire procedure described above is repeated for the alternative function, 
again yielding negligible biases.

\section{Systematic uncertainties}
The uncertainties with the largest impact on the extracted signal yield are those related to the background modeling; since the background parameters are freely floating in the fit, this contribution corresponds to the 
statistical uncertainty in the fitted background. The contribution from the systematic uncertainty in the signal modeling is at least a factor of 3--4 smaller. The dominant sources of systematic uncertainty in the signal
modeling are the jet energy scale, the muon momentum scale, and the muon momentum resolution. Other minor contributions to the signal modeling come from the integrated luminosity, the jet energy resolution, the muon trigger, 
identification, and isolation efficiencies, the \PQb tagging efficiency corrections, and the uncertainties in the renormalization and factorization scales and PDFs.

The uncertainty in the jet energy scale is 2\%, independent of \MLQ, and is propagated by shifting the \Mmuj distributions for the signals by $\pm$2\%. The \Mmuj distribution is also affected by the muon momentum scale, 
with a maximum shift of $\pm$3\%. The uncertainty in the muon momentum resolution translates into an uncertainty in the resolution of \Mmuj and is considered by smearing the reconstructed signal \Mmuj distributions. 
The uncertainty depends on the muon energy and results in a width increase of at most 8\% in the signal \Mmuj distributions.

The total integrated luminosity has an uncertainty of 1.6\%~\cite{CMS-LUM-17-003,CMS-PAS-LUM-17-004,CMS-PAS-LUM-18-002}, 
which is propagated to the normalization of the signal. The uncertainty in the jet energy resolution translates into an uncertainty of 10\% in the resolution of \Mmuj~\cite{CMS:2016lmd}, and is propagated to the 
limits by smearing the reconstructed signal \Mmuj distributions by 10\%. The uncertainties in the muon trigger, identification, and isolation efficiencies amount to about 1\% each and are propagated to the normalization 
of the signal. The uncertainty in the \PQb tagging efficiency corrections depends on the jet \pt, reaching 13\% (20\%) for \PQb (light-flavor) jets at high \pt, and induces anti-correlated variations between the \PQb-tagged 
and non-\PQb-tagged categories, preserving the total signal yield. The uncertainties in the PDFs and in the renormalization and factorization scales also contribute to the uncertainties in the signal model, leading to 
normalization changes in the ranges 2--3\% for both the scales and the PDFs in the case of the LQs with $\PQu\PGm$ coupling, and in the 4--6\% (scales) and 2--4\% (PDFs) ranges for the $\PQb\PGm$ coupling~\cite{greljo2020leptonquarkfusionhadroncolliders}.

\section{Results}\label{sec:results}
We test for the presence of a scalar leptoquark signal with masses between 1.5 and 5\TeV (up to 3.5\TeV for the \LQbmu scenario) and couplings between 0.1 and 2. The lower edge of the search range follows existing constraints 
from pair production searches~\cite{PhysRevD.109.112003,Pair_umu}, whereas the upper bound on $\lambda$ is set to keep the relative LQ width below ${\approx}10\%$, so that the resonant-search approach remains valid.

\begin{figure}[ht!]
   \centering
   \includegraphics[width=0.48\textwidth]{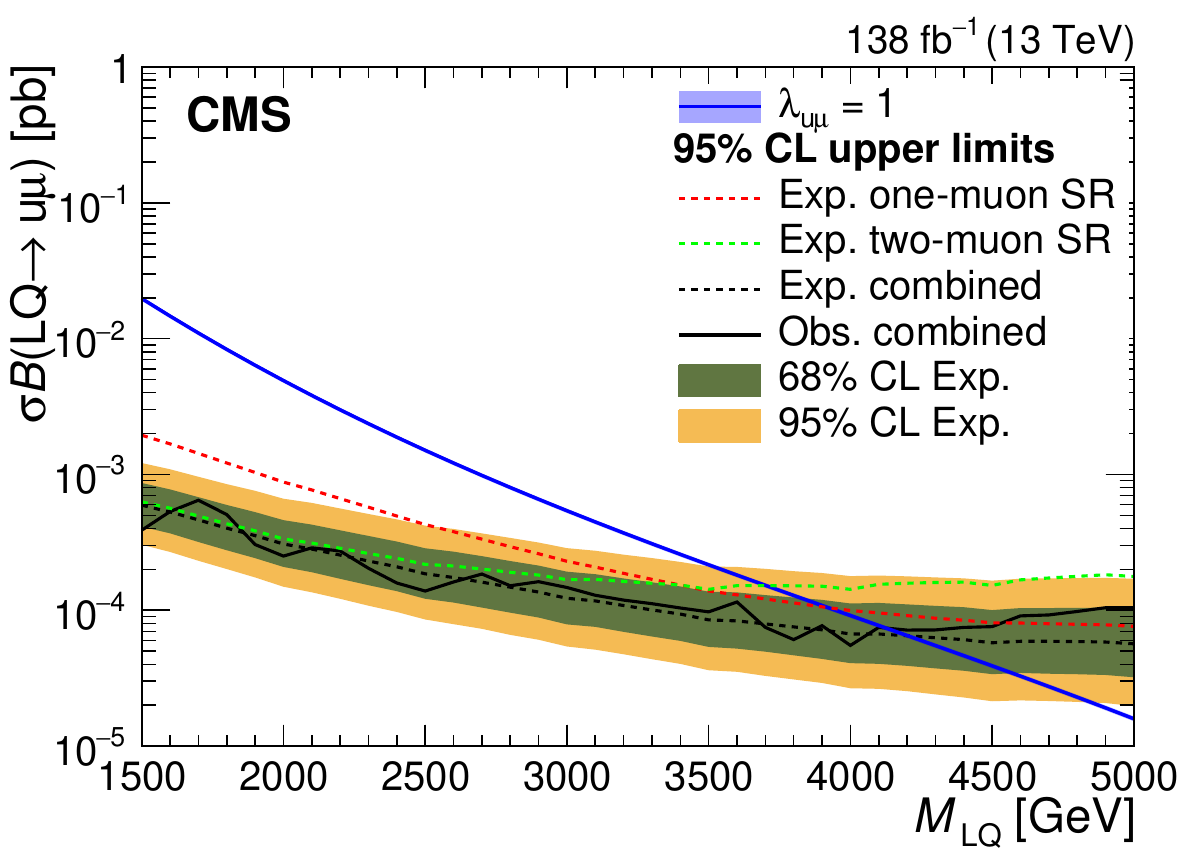}
   \includegraphics[width=0.48\textwidth]{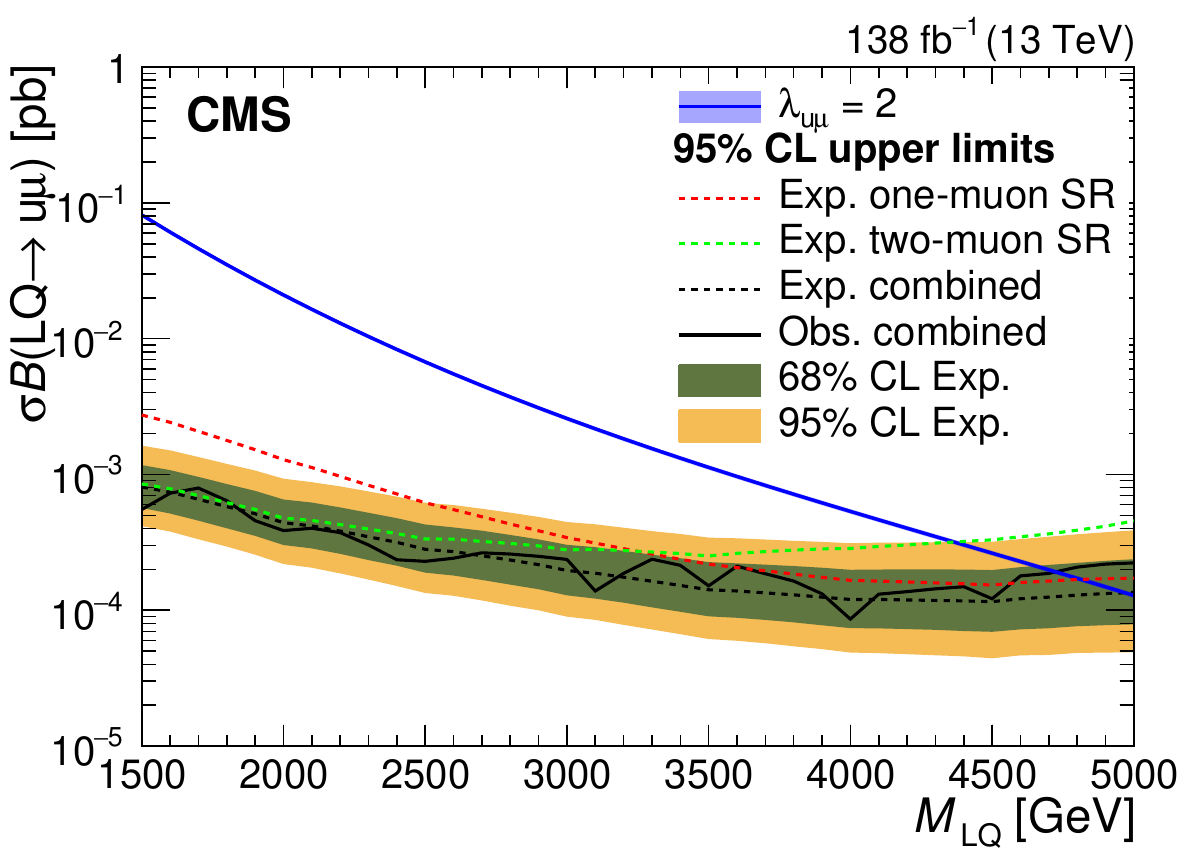}
   \includegraphics[width=0.48\textwidth]{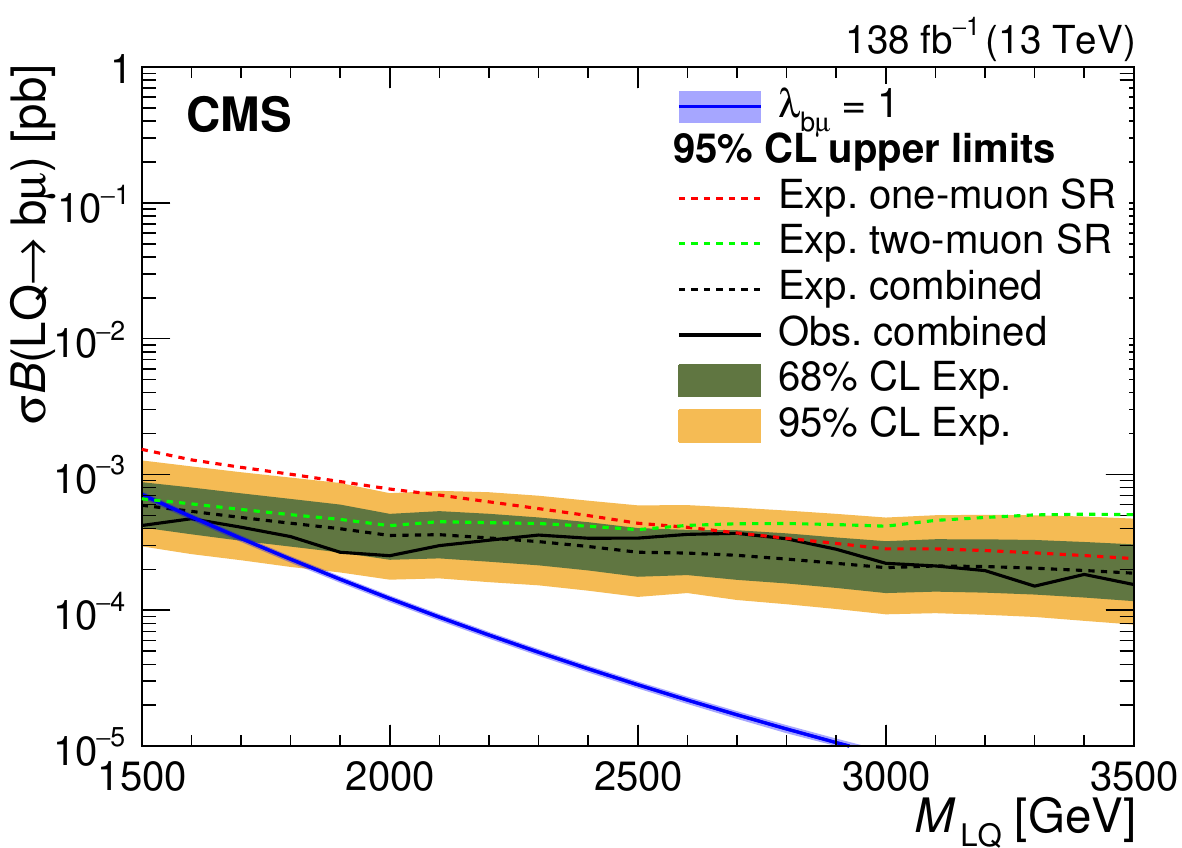} 
   \includegraphics[width=0.48\textwidth]{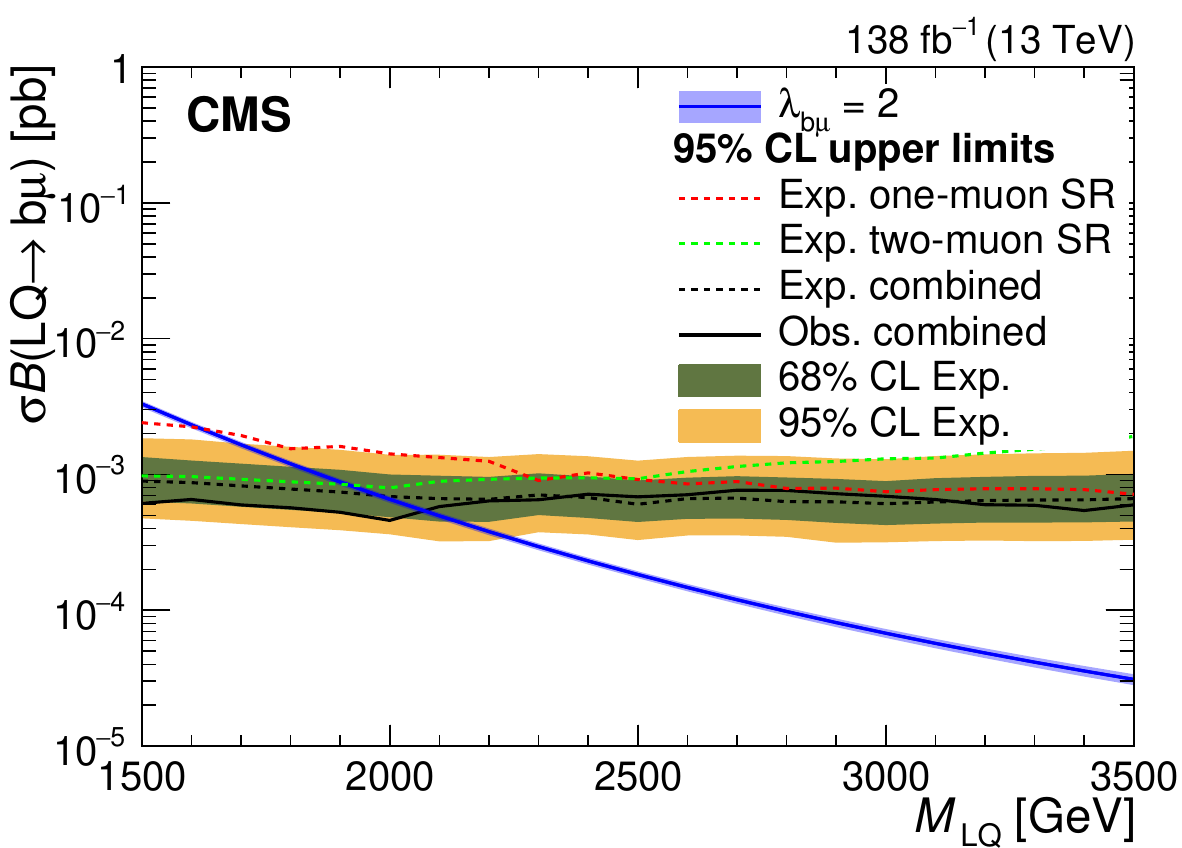} 
   \caption{Expected (dashed black line) and observed (solid black line) 95\% \CL upper limits on $\sigma\mathcal{B}$, as a function of \MLQ, for $\lumu = 1$ (upper left), $\lumu = 2$ (upper right), $\lbmu = 1$ (lower left), 
   and $\lbmu = 2$ (lower right). Uncertainty bands (68\% and 95\% \CL) around the expected limits, as well as the expected limits for the one-muon (red dashed line) and two-muon (green dashed line) SRs, are also shown. 
   The blue lines show the NLO predictions, with the bands around them representing the scale and PDF uncertainties~\cite{greljo2020leptonquarkfusionhadroncolliders}.}
   \label{fig:limits}
\end{figure}

The modified frequentist \CLs criterion~\cite{Junk_1999,ALRead_2002} is used to set upper limits on the signal cross section, computed with the \textsc{Combine} tool~\cite{CMS:2024onh} using the asymptotic approximation 
of the test statistic~\cite{Cowan_2011}. Upper limits at 95\% confidence level (\CL) are set on the product of the LQ production cross section and branching fraction, $\sigma\mathcal{B}$, where $\mathcal{B}$ is the 
branching fraction to the muon-quark final state, for the different \MLQ and $\lambda$ values for the \LQumu and \LQbmu scenarios. The 95\% \CL upper limits on $\sigma\mathcal{B}$ as a function of \MLQ for 
$\lumu = 1$ and 2 and $\lbmu = 1$ and 2 are shown in Fig.~\ref{fig:limits}, together with the NLO predictions. For the \LQbmu scenario, the two-muon SR provides the best expected sensitivity at low masses, 
while the one-muon SR dominates at higher masses.

The cross section limits are computed assuming only the lepton-induced single LQ production process, as the analysis was specifically optimized for this production mechanism. For simplicity, results from other production 
methods, in particular gluon-induced single production, are not included. This choice yields conservative (weaker) upper limits, by approximately 10--15\%.
As a cross-check, we generated pseudo-data with all signal production processes included and then performed the fit using only the lepton-induced signal shape. The resulting bias in the signal extraction, 
evaluated as $(\text{fitted signal} - \text{injected signal}) / \sigma_\text{fit}$, is found to be below half a standard deviation, and typically much smaller, across all tested masses and couplings, 
confirming that the analysis remains sensitive to the signal even when only the lepton-induced production is considered.

\begin{figure}[ht!]
   \centering
   \includegraphics[width=0.48\textwidth]{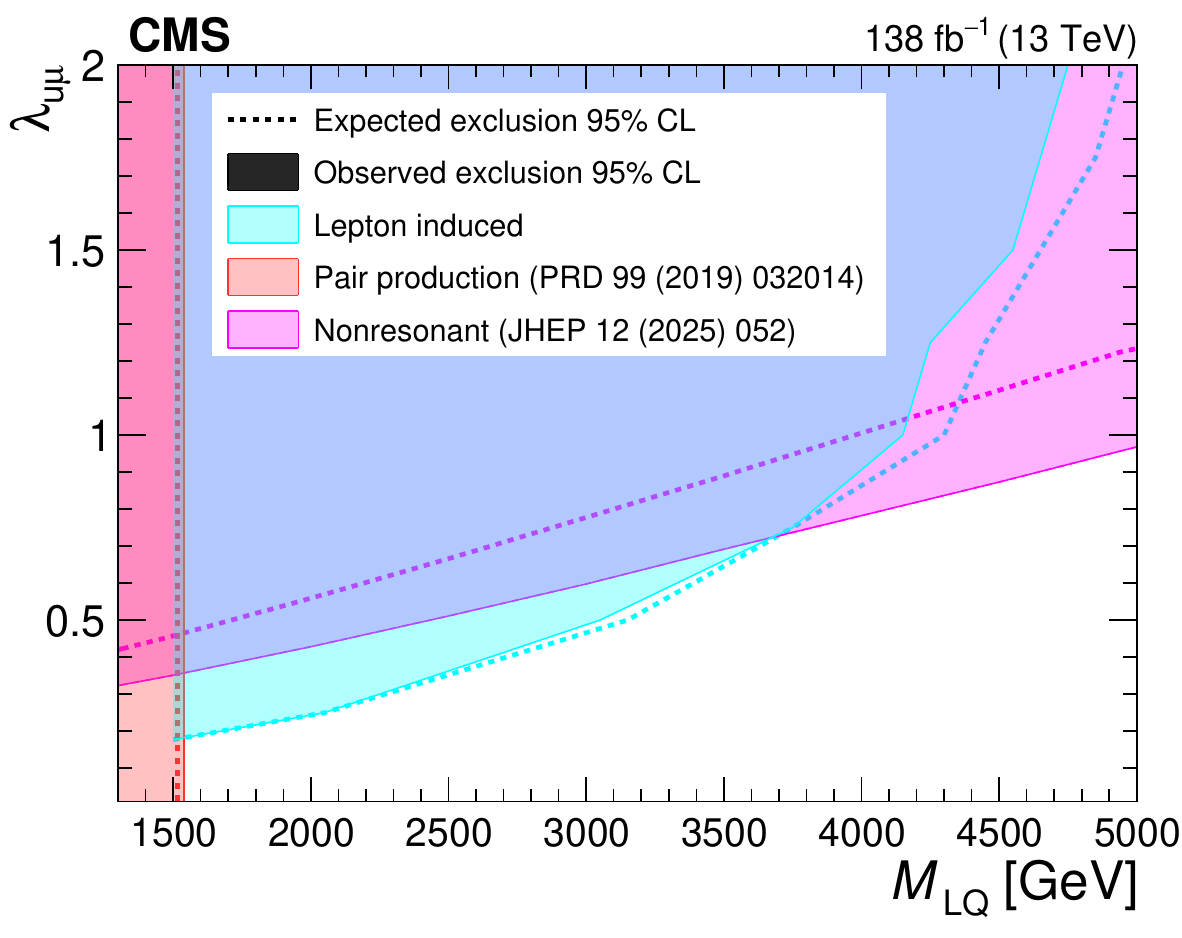}
   \includegraphics[width=0.48\textwidth]{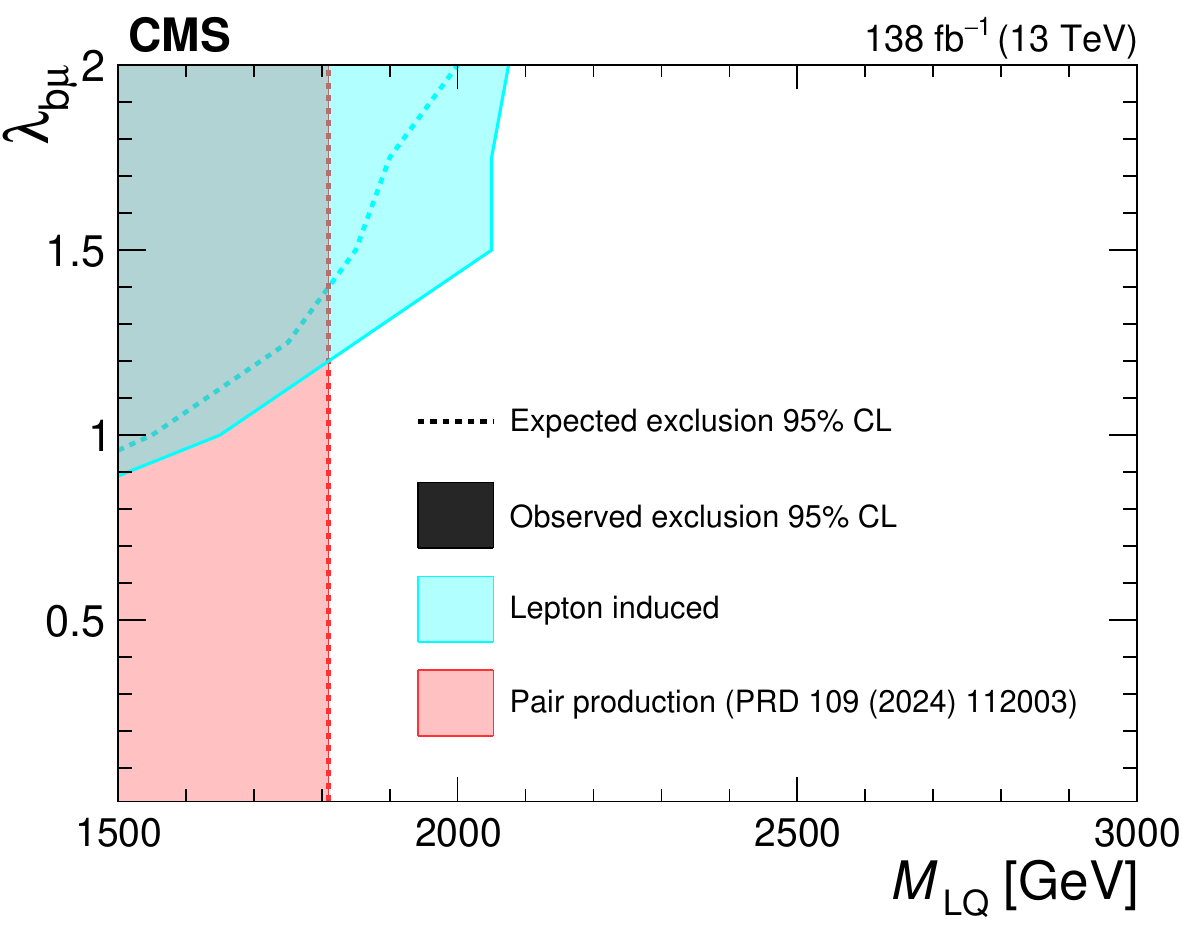}
   \caption{95\% \CL exclusion regions in the (\MLQ, $\lambda$) plane, for the \LQumu (left) and \LQbmu (right) scenarios. For each production mode, shown in a different color, the dashed contour indicates the expected 
   exclusion and the shaded area the observed one, as illustrated by the black keys in the legend. The cyan area shows the region excluded by the lepton-induced single LQ production considered in this analysis. 
   The red and magenta areas show the regions excluded by previous CMS searches targeting the pair production~\cite{PhysRevD.109.112003,Pair_umu} and the nonresonant production~\cite{CMS:2025iix} modes, respectively. 
   The nonresonant production constraint is available only for the \LQumu scenario.}
   \label{fig:MassvsCoup_umu}
\end{figure}

The results are also presented in the (\MLQ, $\lambda$) plane in Fig.~\ref{fig:MassvsCoup_umu} for the \LQumu and \LQbmu hypotheses, together with the exclusion contours from previous CMS LQ analyses targeting the pair 
production mode~\cite{Pair_umu,PhysRevD.109.112003}, which are largely independent of the coupling value, and the nonresonant single production mode~\cite{CMS:2025iix}. For each production mode, the shaded area (hatched contour) 
represents the observed (expected) region of the plane corresponding to the excluded signal hypotheses. The different shapes of the exclusion contours reflect the underlying production mechanisms: pair production is dominated by 
QCD interactions and depends mainly on \MLQ, while single-production cross sections scale with $\lambda^2$. As a result, the pair production exclusion is largely independent of $\lambda$ and appears as an approximately vertical 
boundary in the (\MLQ, $\lambda$) plane. This search benefits from the resonant topology of the lepton-induced production and gains sensitivity relative to pair production searches at large $\lambda$ and high \MLQ, 
where the pair production cross section is strongly suppressed by phase space. At the same time, with increasing $\lambda$ and \MLQ, the resonant production cross section decreases more rapidly than the nonresonant contribution, 
reducing the relative sensitivity of this search in that regime. The analysis presented here extends the CMS coverage to previously unconstrained regions of the (\MLQ, $\lambda$) plane, specifically, $1.5 < \MLQ < 3.6\TeV$ 
and $0.2 < \lambda < 0.6$ for the \LQumu scenario, and $\MLQ > 1.8\TeV$ and $\lambda > 1$ for the \LQbmu scenario.

\section{Summary}
A search for scalar leptoquarks (LQs) coupled to muons, using proton-proton collision data at $\sqrt{s} = 13\TeV$ collected by the CMS experiment at the LHC and corresponding to an integrated luminosity of 138\fbinv, 
has been presented. The final state includes one or two muons (\PGm) and a jet that can come from the hadronization of either a light quark (\PQu) or a \PQb quark. The analysis extends the search for LQs coupled to 
muons by exploiting a new production mechanism that leverages the charged-lepton content of protons arising from quantum fluctuations. This mechanism enables the study of lepton-induced processes and resonant single 
LQ production and decay at the LHC.

The data are found to be in agreement with the background predictions. Upper limits on the product of the LQ production cross section and branching fraction to the muon-quark final state are derived.
These limits exclude LQs with masses between 1.5 and 3.6\TeV, and couplings between 0.2 and 0.6 for the \LQumu scenario; and masses above 1.8\TeV, and couplings above 1 for the \LQbmu scenario. 
These results extend the mass and coupling ranges probed by previous searches in other production modes.

\begin{acknowledgments}
We congratulate our colleagues in the CERN accelerator departments for the excellent performance of the LHC and thank the technical and administrative staffs at CERN and at other CMS institutes for their contributions to the success of the CMS effort. In addition, we gratefully acknowledge the computing centers and personnel of the Worldwide LHC Computing Grid and other centers for delivering so effectively the computing infrastructure essential to our analyses. Finally, we acknowledge the enduring support for the construction and operation of the LHC, the CMS detector, and the supporting computing infrastructure provided by the following funding agencies: SC (Armenia), BMFWF and FWF (Austria); FNRS and FWO (Belgium); CNPq, CAPES, FAPERJ, FAPERGS, and FAPESP (Brazil); MES and BNSF (Bulgaria); CERN; CAS, MoST, and NSFC (China); MINCIENCIAS (Colombia); MSES and CSF (Croatia); RIF (Cyprus); SENESCYT (Ecuador); ERC PRG and PSG, TARISTU24-TK10 and MoER TK202 (Estonia); Academy of Finland, MEC, and HIP (Finland); CEA and CNRS/IN2P3 (France); SRNSF (Georgia); BMFTR, DFG, and HGF (Germany); GSRI (Greece); MATE and NKFIH (Hungary); DAE and DST (India); IPM (Iran); SFI (Ireland); INFN (Italy); MSIT and NRF (Republic of Korea); MES (Latvia); LMTLT (Lithuania); MOE and UM (Malaysia); BUAP, CINVESTAV, CONACYT, LNS, SEP, and UASLP-FAI (Mexico); MOS (Montenegro); MBIE (New Zealand); PAEC (Pakistan); MSHE, NSC, and NAWA (Poland); FCT (Portugal); MESTD (Serbia); MICIU/AEI and PCTI (Spain); MOSTR (Sri Lanka); Swiss Funding Agencies (Switzerland); MST (Taipei); MHESI (Thailand); TUBITAK and TENMAK (T\"{u}rkiye); NASU (Ukraine); STFC (United Kingdom); DOE and NSF (USA).

\begin{sloppypar}
\setlength\emergencystretch{\hsize}
\hyphenation{Rachada-pisek} Individuals have received support from the Marie-Curie program and the European Research Council and Horizon 2020 Grant, contract Nos.\ 675440, 724704, 752730, 758316, 765710, 824093, 101115353, 101002207, 101001205, and COST Action CA16108 (European Union); the Leventis Foundation; the Alfred P.\ Sloan Foundation; the Alexander von Humboldt Foundation; the Science Committee, project no. 22rl-037 (Armenia); the Fonds pour la Formation \`a la Recherche dans l'Industrie et dans l'Agriculture (FRIA) and Fonds voor Wetenschappelijk Onderzoek contract No. 1228724N (Belgium); the Beijing Municipal Science \& Technology Commission, No. Z191100007219010, the Fundamental Research Funds for the Central Universities, the Ministry of Science and Technology of China under Grant No. 2023YFA1605804, the Natural Science Foundation of China under Grant No. 12535004, and USTC Research Funds of the Double First-Class Initiative No.\ YD2030002017 (China); the Ministry of Education, Youth and Sports (MEYS) of the Czech Republic; the Shota Rustaveli National Science Foundation (Georgia); the Deutsche Forschungsgemeinschaft (DFG), among others, under Germany's Excellence Strategy -- EXC 2121 ``Quantum Universe" -- 390833306, and under project number 400140256 - GRK2497; the Hellenic Foundation for Research and Innovation (HFRI), Project Number 2288 (Greece); the Hungarian Academy of Sciences, the New National Excellence Program - \'UNKP, the NKFIH research grants K 131991, K 138136, K 143460, K 143477, K 147557, K 146913, K 146914, K 147048, TKP2021-NKTA-64, and 2025-1.1.5-NEMZ\_KI-2025-00004, and MATE KKP and KKPCs Research Excellence and Flagship Research Groups grants (Hungary); the Council of Science and Industrial Research, India; ICSC -- National Research Center for High Performance Computing, Big Data and Quantum Computing, FAIR -- Future Artificial Intelligence Research, and CUP I53D23001070006 (Mission 4 Component 1), funded by the NextGenerationEU program, the Italian Ministry of University and Research (MUR) under Bando PRIN 2022 -- CUP I53C24002390006, PRIN PRIMULA 2022RBYK7T (Italy); the Latvian Council of Science; the Ministry of Science and Higher Education, project no. 2022/WK/14, and the National Science Center, contracts Opus 2021/41/B/ST2/01369, 2021/43/B/ST2/01552, 2023/49/B/ST2/03273, and the NAWA contract BPN/PPO/2021/1/00011 (Poland); the Funda\c{c}\~ao para a Ci\^encia e a Tecnologia (Portugal); the National Priorities Research Program by Qatar National Research Fund; MICIU/AEI/10.13039/501100011033, ERDF/EU, ``European Union NextGenerationEU/PRTR", projects PID2022-142604OB-C21, PID2022-139519OB-C21, PID2023-147706NB-I00, PID2023-148896NB-I00, PID2023-146983NB-I00, PID2023-147115NB-I00, PID2023-148418NB-C41, PID2023-148418NB-C42, PID2023-148418NB-C43, PID2023-148418NB-C44, PID2024-158190NB-C22, RYC2021-033305-I, RYC2024-048719-I, CNS2023-144781, CNS2024-154769 and Plan de Ciencia, Tecnolog{\'i}a e Innovaci{\'o}n de Asturias, Spain; the Chulalongkorn Academic into Its 2nd Century Project Advancement Project, the National Science, Research and Innovation Fund program IND\_FF\_68\_369\_2300\_097, and the Program Management Unit for Human Resources \& Institutional Development, Research and Innovation, grant B39G680009 (Thailand); the Eric \& Wendy Schmidt Fund for Strategic Innovation through the CERN Next Generation Triggers project under grant agreement number SIF-2023-004; the Kavli Foundation; the Nvidia Corporation; the SuperMicro Corporation; the Welch Foundation, contract C-1845; and the Weston Havens Foundation (USA).
\end{sloppypar}
\end{acknowledgments}\section*{Data availability} Release and preservation of data used by the CMS Collaboration as the basis for publications is guided by the  \href{https://doi.org/10.7483/OPENDATA.CMS.1BNU.8V1W}{CMS data preservation, re-use and open access policy}.

\bibliography{auto_generated}

\cleardoublepage \appendix\section{The CMS Collaboration \label{app:collab}}\begin{sloppypar}\hyphenpenalty=5000\widowpenalty=500\clubpenalty=5000\input{EXO-24-005-public-authorlist.tex}\end{sloppypar}
\end{document}

%% file: EXO-24-005-public-authorlist.tex
\cmsinstitute{Yerevan Physics Institute, Yerevan, Armenia}
{\tolerance=6000
A.~Gevorgyan\cmsorcid{0000-0003-2751-9489}, A.~Hayrapetyan, V.~Makarenko\cmsorcid{0000-0002-8406-8605}, A.~Tumasyan\cmsAuthorMark{1}\cmsorcid{0009-0000-0684-6742}
\par}
\cmsinstitute{Marietta Blau Institute for Particle Physics, Vienna, Austria}
{\tolerance=6000
W.~Adam\cmsorcid{0000-0001-9099-4341}, L.~Benato\cmsorcid{0000-0001-5135-7489}, T.~Bergauer\cmsorcid{0000-0002-5786-0293}, M.~Dragicevic\cmsorcid{0000-0003-1967-6783}, S.~Gundacker\cmsorcid{0000-0003-2087-3266}, A.K.~Guven\cmsorcid{0009-0004-5670-5138}, P.S.~Hussain\cmsorcid{0000-0002-4825-5278}, M.~Jeitler\cmsAuthorMark{2}\cmsorcid{0000-0002-5141-9560}, N.~Krammer\cmsorcid{0000-0002-0548-0985}, A.~Li\cmsorcid{0000-0002-4547-116X}, D.~Liko\cmsorcid{0000-0002-3380-473X}, M.~Matthewman, J.~Schieck\cmsAuthorMark{2}\cmsorcid{0000-0002-1058-8093}, R.~Sch\"{o}fbeck\cmsAuthorMark{2}\cmsorcid{0000-0002-2332-8784}, M.~Shooshtari\cmsorcid{0009-0004-8882-4887}, M.~Sonawane\cmsorcid{0000-0003-0510-7010}, N.~Van~Den~Bossche\cmsorcid{0000-0003-2973-4991}, W.~Waltenberger\cmsorcid{0000-0002-6215-7228}, C.E.~Wulz\cmsAuthorMark{2}\cmsorcid{0000-0001-9226-5812}
% ...
\par}
$^{1}$Also at Yerevan State University, Yerevan, Armenia\\
$^{2}$Also at Technische Universit\"{a}t Wien, Vienna, Austria\\
% ...
